\documentclass[journal]{IEEEtran}
\usepackage[colorlinks,citecolor=red,urlcolor=blue,bookmarks=false,hypertexnames=true]{hyperref}
\usepackage{graphicx}
\usepackage{amsmath,amsfonts}
\usepackage{algorithmic}
\usepackage{algorithm}
\usepackage{array}
\usepackage{textcomp}
\usepackage{subcaption}
\usepackage{stfloats}
\usepackage{url}
\usepackage{verbatim}
\usepackage{graphicx}
\usepackage{cite}
\usepackage{multirow}
\usepackage{optidef}
\usepackage{subcaption}

\begin{document}
\title{Communication Security and Sensing Privacy in FMCW-Based ISAC Through Signal Modulation}

\author{Murat~Temiz,~\IEEEmembership{Member,~IEEE,}
        Christos~Masouros,~\IEEEmembership{Fellow,~IEEE}
\thanks{M. Temiz and C. Masouros declare a relevant patent application: United Kingdom Patent Application No. 2521171.5. The authors are with the Department of Electronic and Electrical Engineering, University College London, London, WC1E 7JE, United Kingdom. M. Temiz is also with METU, Ankara, Türkiye.}
}

\markboth{Journal of \LaTeX\ Class Files,~Vol.~14, No.~8, August~2015}%
{Shell \MakeLowercase{\textit{et al.}}: Bare Demo of IEEEtran.cls for IEEE Journals}

\maketitle

\begin{abstract}
This study proposes a novel radar-centric signaling design and architecture for secure integrated sensing and communication (ISAC) systems. The proposed framework is designed to provide robust physical layer security for data transmission while simultaneously enhancing sensing privacy. It employs index modulation and phase coding over frequency-modulated continuous-wave radar (FMCW) chirps, where index modulation (IM) provides an outer layer of data security, and we explicitly design the phase coding (PC) to perturb the resulting signal's ambiguity function (AF) to enhance sensing privacy. This design reduces the risk of unauthorized surveillance by rendering target velocity estimation practically infeasible for unauthorized passive sensing hardware (i.e., a sensing eavesdropper, S-Eve) and significantly impairing its range estimation capabilities. Furthermore, this study also presents the transmitter and receiver architectures required for effective modulation and demodulation of the proposed ISAC signaling and for performing sensing at the legitimate sensing hardware. Simulation results show that the proposed approach achieves high data throughput while enhancing communication security and sensing privacy.
\end{abstract}


\IEEEpeerreviewmaketitle

\section{Introduction}

\IEEEPARstart{T}he integration of sensing and communication systems introduces new security and privacy risks stemming from the dual use of wireless waveforms for both data transmission and environmental sensing. In particular, communication security becomes more vulnerable because signals designed for active sensing inherently carry information, allowing a sensing target to function as a potential communication eavesdropper \cite{Su2023}. On the other hand, ISAC signals might be used by malicious devices for passive sensing, whereby adversarial devices can exploit existing integrated communication and sensing (ISAC) signals to infer environmental and behavioral information without emitting detectable transmissions \cite{WeiISAC6G2022}. Such passive sensing capabilities significantly increase sensing-privacy threats, as individuals may be monitored or profiled without awareness or consent. These risks are especially severe in safety-critical scenarios, where passive sensing by unauthorized devices can lead to covert surveillance and sensitive information leakage.

While encryption at the upper layers of communication systems enhances data security, it is essential to address physical layer security issues to minimize information leakage. This is because encryption does not conceal physical-layer signal characteristics, such as the power spectrum and waveform structure, which may still be exploited for jamming, interception, or signal analysis \cite{shakiba2021physical}. Furthermore, encryption does not address sensing privacy, since sensing-related information is often embedded in the transmitted waveform itself rather than in the encrypted payload data. On the other hand, physical layer security techniques facilitate secure signal transmission, including data frames and headers, thereby improving overall communication security and sensing privacy \cite{MaSensing2025}. Furthermore, unlike traditional encryption algorithms, physical-layer security techniques generally maintain lower computational complexity by avoiding resource-intensive cryptographic processing. \cite{YangSafeguarding2015}.

Physical-layer data security has recently attracted significant attention in communication and ISAC systems toward 6G networks \cite{OnurSecureISAC2023, mao2023security, zhu2024enabling}. For instance, one study investigates MIMO radar beamforming designs that balance secrecy rate maximization with accurate target detection beampatterns \cite{DeligiannisSecrecy2018}. Addressing scenarios in which sensing targets act as potential eavesdroppers, artificial-noise beamforming toward malicious targets is proposed to degrade their eavesdropping capability without affecting legitimate communication users \cite{su2020secure, ren2023robust}. In a related work, the sensing functionality is leveraged to estimate eavesdropper locations, enabling more accurate null steering towards them \cite{SuSensingISAC2024}. Another study enhances physical-layer security by exploiting constructive interference, transforming multi-user interference into a mechanism that strengthens legitimate signals while suppressing eavesdroppers \cite{su2025secure}. Moreover, a two-stage transmission protocol for secure ISAC is proposed in \cite{CaoSecureISAC2025}, where an optimization problem is formulated to suppress eavesdropping by jointly controlling sensing and communication performance.

On the other hand, passive sensing-based methods exploit the available radio frequency (RF) signals in the environment for sensing human activities, vehicles, UAVs, and other targets \cite{TaylorPassiveRadar2025, SavvidouPassiveRadar2024}. The sources of the RF signals can be communication, broadcasting, other sensing, or ISAC systems, and these systems cannot know if their signals are being used for passive sensing. Passive sensing raises concerns about privacy because it allows the detection of individuals in homes, businesses, or restricted areas without the need to transmit radio frequency (RF) signals \cite{SunWiFi2021}. Additionally, machine learning applications can gather a significant amount of information from data obtained through passive sensing, which may include target classification and recognition \cite{NirmalDeepLearning2021}.

\begin{figure}
\centering\includegraphics[width=1\linewidth]{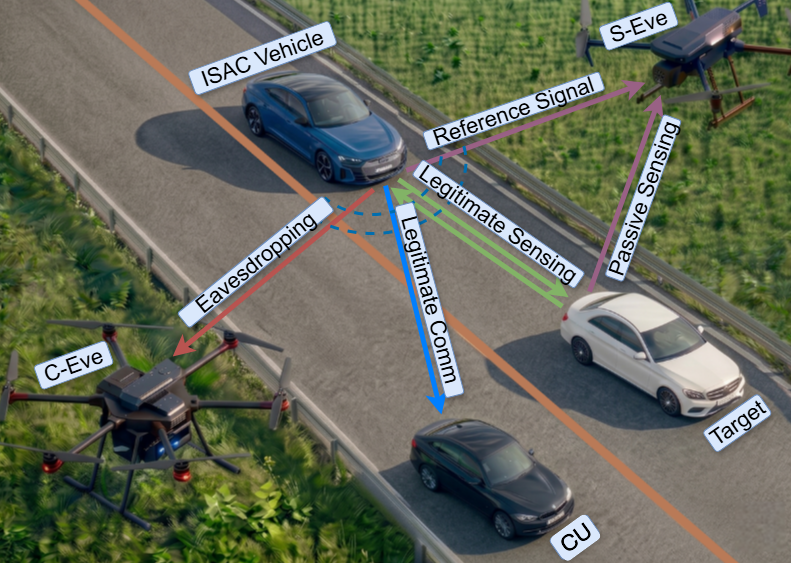}
    \caption{A typical ISAC scenario, where the ISAC transceiver transmits ISAC signals to communicate with the CU and perform sensing using the same signals. Meanwhile, the S-Eve aims to exploit the ISAC signals for unauthorized sensing, and the C-Eve attempts to eavesdrop on the communication.}
    \label{fig:scenario}
\end{figure}

Various methods have been recently developed against passive sensing, mainly for communication-only systems. For instance, generating specific signals is proposed to jam communication signals towards specific directions to blind adversary passive sensing devices \cite{nguyen2014real, schulz2017massive}. Other studies aim to incapacitate the RF sensing of the eavesdropper by adding random amplitude, delay, and Doppler shift variations to the transmitted signals \cite{YaoRFShield2024, yao2018aegis, argyriou2023obfuscation, MahmudBeam2021}. These variations can also mimic micro-Doppler signatures of human activities to confuse the sensing eavesdropper. Moreover, Cominelli et al. proposed techniques to reduce the localization ability of adversary passive sensors  \cite{cominelli2022antisense, CominelliWiFiCSI2021, GhiroLoca2022} by adding pseudo-random noise that can be filtered out at the legitimate receiver. Scheduling the distributed antennas is also considered to reduce the sensing ability of adversarial passive sensing devices \cite{hernandez2023scheduled}.   

Sensing privacy has been recently considered in ISAC systems. For instance, beamforming artificial noise is considered to reduce the sensing capabilities of a sensing eavesdropper, which is also one of the legitimate communication users in the network \cite{ZouSecureSensing2024}. Han et al. have explored ambiguity function (AF) design for orthogonal frequency-division multiplexing (OFDM) signaling in ISAC systems to improve the sensing privacy, where OFDM signals are designed to have high sidelobes to degrade the sensing capability of sensing eavesdroppers \cite{HanISACSecure2025}. AF design based on the Kullback-Leibler divergence is also proposed to intentionally degrade the detection probability of unauthorized eavesdroppers while maintaining the performance of legitimate sensing and communication\cite{du2025securing}. Another study proposes a reconfigurable intelligent surface (RIS)-aided framework that jointly optimizes beamforming and phase shifts to intentionally degrade the sensing signal-to-interference-plus-noise ratio (SINR) of eavesdroppers while maintaining performance for users \cite{magbool2025hiding}. A near-field ISAC technique is also proposed for sensing privacy, where the strategic illumination of environmental scatterers is utilized to create deceptive clutter that masks true target parameters from unauthorized observers without requiring their specific channel information \cite{chen2025sensing}. Another study introduces the security and privacy-preserving network (SPPN) framework, which employs AI-driven decision-makers, friendly jammers, and RIS to protect sensitive data while enhancing overall system performance \cite{qu2024privacy}. Moreover, recent surveys have comprehensively reviewed communication security vulnerabilities and sensing privacy issues introduced by the integration of sensing and communications, and provided possible countermeasures to protect both data transmission and sensing privacy \cite{han2025next, qu2026secure2026}.

In this study, we have proposed a radar-centric secure ISAC signaling design. Previously proposed radar-centric ISAC waveforms and systems add communication capacity to the existing radar waveforms by manipulating them according to communication data \cite{HuangMajorCom2020, TemizISAC2023, TemizISACCon2023, XuHybrid2023, MaFrac2021, HanRadCom2023}. For instance, the combination of frequency selection and spatial antenna selection is utilized to modulate data within a pulsed multi-antenna radar system \cite{HuangMajorCom2020}. Moreover, a combination of phase modulation, antenna selection, and frequency selection of pulses is used to transmit data within radar pulses \cite{XuHybrid2023}. Another study utilized carrier frequency selection, antenna selection, and phase modulation with FMCW signals to enable data transmission within radar systems \cite{MaFrac2021}.

Existing works on secure and privacy-preserving ISAC mainly address beamforming \cite{ZouSecureSensing2024}, artificial noise generation \cite{su2020secure, ren2023robust}, or ambiguity-function shaping for OFDM-based waveforms \cite{HanISACSecure2025, du2025securing}. In contrast, this study utilizes the selection of center frequency and bandwidth of frequency modulated continuous wave (FMCW) chirps, along with phase coding and codebook design, to form a secure ISAC signaling. The security-enhanced modulation is based on a codebook that ensures that the transmitted signals cannot be easily used for passive sensing by an adversary's passive sensing hardware. Moreover, due to the proposed two-layer data modulation scheme and codebook design, the physical layer security is also improved such that the data transmission cannot be intercepted by an eavesdropper. 

The main contributions of this study are as follows. A radar-centric secure signaling and codebook design framework is proposed for ISAC systems to enhance physical-layer data security and sensing privacy. In addition, transmitter and receiver architectures and the associated algorithms are developed for the proposed secure ISAC signaling, and their performance is evaluated. Finally, the interplay among sensing privacy, data rate, and communication security is analyzed.

\emph{Notation:} Bold lowercase and uppercase letters denote vectors and matrices, respectively. The operators $(\cdot)^{T}$, $(\cdot)^{*}$, $|\cdot|$, $\lfloor \cdot \rfloor$, $\lceil \cdot \rceil$, $\odot$, and $\oslash$ denote the transpose, complex conjugate, absolute value, floor, ceiling, element-wise product, and element-wise division, respectively. The norms $\|\cdot\|_2$ and $\|\cdot\|_0$ denote the Euclidean norm and the number of nonzero entries of a vector, respectively.

\section{System Model}

A typical ISAC scenario is presented in Fig.~\ref{fig:scenario}, where a base station (BS) transmits ISAC signals to communicate with the legitimate communication user (CU) while also performing monostatic sensing by processing the echoes of ISAC signals through its sensing receiver hardware. However, an adversary device, acting as a sensing eavesdropper (S-Eve), may aim to exploit these ISAC signals to perform passive sensing. Meanwhile, a communication eavesdropper (C-Eve) may attempt to eavesdrop on the communication between the ISAC BS and the CU. Such unauthorized hardware can exploit the ISAC signals, which can cause data security and privacy issues. Thus, in the following, a novel signaling and codebook design technique is proposed to improve the security of the ISAC system against adversary sensing and interception of the communication data. 

\begin{figure*}
\centering\includegraphics[width=0.8\linewidth]{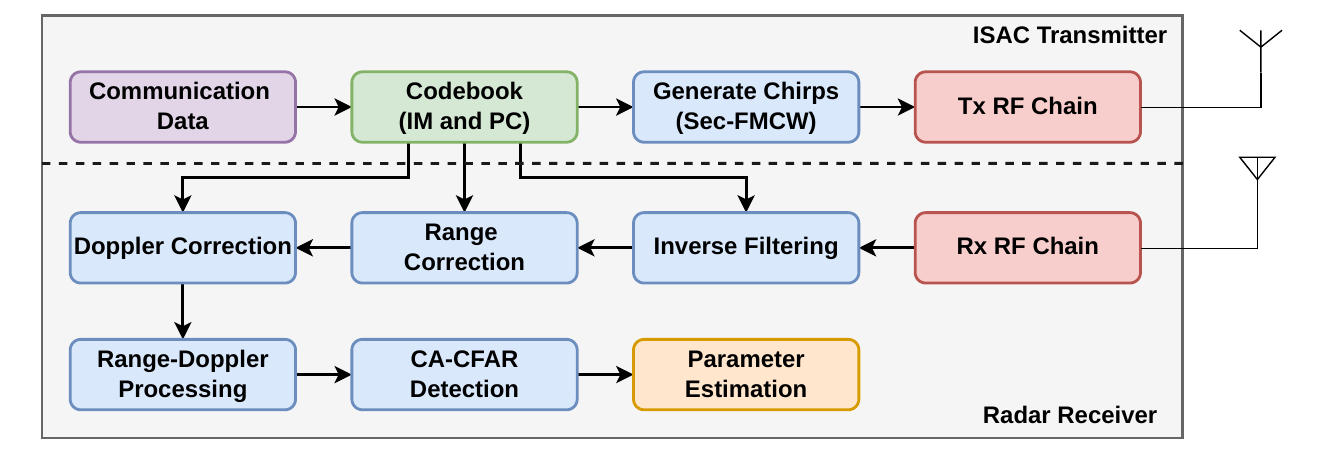}
    \caption{The proposed ISAC transceiver that transmits ISAC signals for communications and sensing, and receives and processes the signals for target detection and parameter estimation.}
\label{fig:Isac_transceiever}
\end{figure*}

\begin{figure*}
\centering\includegraphics[width=1\linewidth]{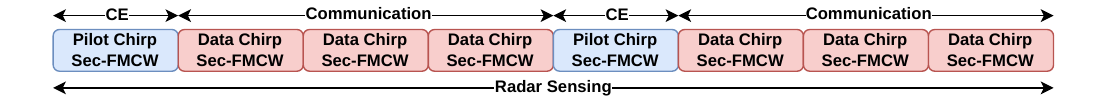}
    \caption{Transmission of Sec-FMCW signaling, including pilot chirps for channel estimation (CE).}
\label{fig:transmission_frame}
\end{figure*}

\subsection{ISAC Transmitter Architecture}

For the secure ISAC signaling, the ISAC transceiver architecture shown in Fig.~\ref{fig:Isac_transceiever} is proposed, which is comprised of an ISAC transmitter that generates the secure ISAC signals and a radar receiver that performs the detection and estimation of target parameters.  The proposed secure signaling is digitally generated as baseband chirps as shown in Fig.~\ref{fig:Isac_transceiever}, and then upconverted to RF signals and transmitted by the antenna. For the communication data, the corresponding codeword from the codebook is selected, and the corresponding Sec-FMCW chirps are generated. The pilot chirps are placed between a certain number of Sec-FMCW chirps to allow the communication receiver to perform channel estimation and channel equalization, as shown in Fig~\ref{fig:transmission_frame}. After that, the chirps are converted via a transmitter (Tx) radio frequency (RF) chain, and these RF signals are transmitted by the antenna.

\subsection{Signal Model}

FMCW signals are widely employed in short-range radar systems due to having very low sidelobes for delay and Doppler estimations in their AFs. Moreover, because of their simple and periodic signal structure, they can be generated and processed by low-complexity transmitter and receiver hardware.  The proposed secure radar-centric ISAC signaling utilizes index-modulation and phase codes within FMCW chirps to transmit data while performing sensing. In doing so, the phase codes are specifically constructed to create artificial sidelobes in the desired location of the AFs of the chirps to prevent the signals from being used for passive sensing by an adversary device. 

The transmitted frames in vertical polarization (V-pol) and horizontal polarization (H-pol) are denoted by vectors $\mathbf{x}^V$ and $\mathbf{x}^H$, respectively. Each of the transmitted frames in V-pol and H-pol consists of $I$ chirps, including pilot and data chirps, obtained by concatenating the chirps as,
\begin{align}
    &\mathbf{x}^V = \left[\mathbf{x}_1^V, \dots, \mathbf{x}_i^V, \dots, \mathbf{x}_I^V\right]  \notag\\ 
    &\mathbf{x}^H = \left[\mathbf{x}_1^H, \dots, \mathbf{x}_i^H, \dots, \mathbf{x}_I^H\right],
\end{align}
where vectors $\mathbf{x}_i^V \in \mathbb{C}^{1 \times T}$ and $\mathbf{x}_i^H \in \mathbb{C}^{1 \times T}$ denote the $i$-th chirps transmitted in V-pol and H-pol, respectively, and they are given by
\begin{align}
    &\mathbf{x}_i^V = \left[x^V_1(1), \dots, x^V_i(t), \dots, x^V_i(T)  \right]  \notag\\ 
    &\mathbf{x}_i^H = \left[x^H_i(1), \dots, x^H_i(t), \dots, x^H_i(T)  \right],
\end{align} where $x^V_i(t)$ and $x^H_i(t)$ denote the complex-valued samples at time $t$ with $1\leq t \leq T$. The $i$th complex-valued FMCW chirp transmitted in V-pol or H-pol ($x^V_i(t)$  or $x^H_i(t)$ ) can be given by,
\begin{equation}
x_i(t)=\cos\left(\theta_i(t)+\phi_{re}^l(t)\right)+j\sin\left({\theta_i(t)+\phi_{im}^l(t)}\right), \label{fmcw_cis}
\end{equation}
where $\theta_i(t)$ denotes the instantaneous frequency of the chirp at time $t$, and $\phi_{re}^l(t)$ and $\phi_{im}^l(t)$ denote the phases of the real and imaginary part of the baseband signal for each $l$th segment of the chirp consisting of $L$ segments (chips), such that $l = 1, 2, \dots, L$. 

\begin{figure}
    \centering
    \begin{subfigure}{0.5\textwidth}
        \centering  \includegraphics[width=\textwidth]{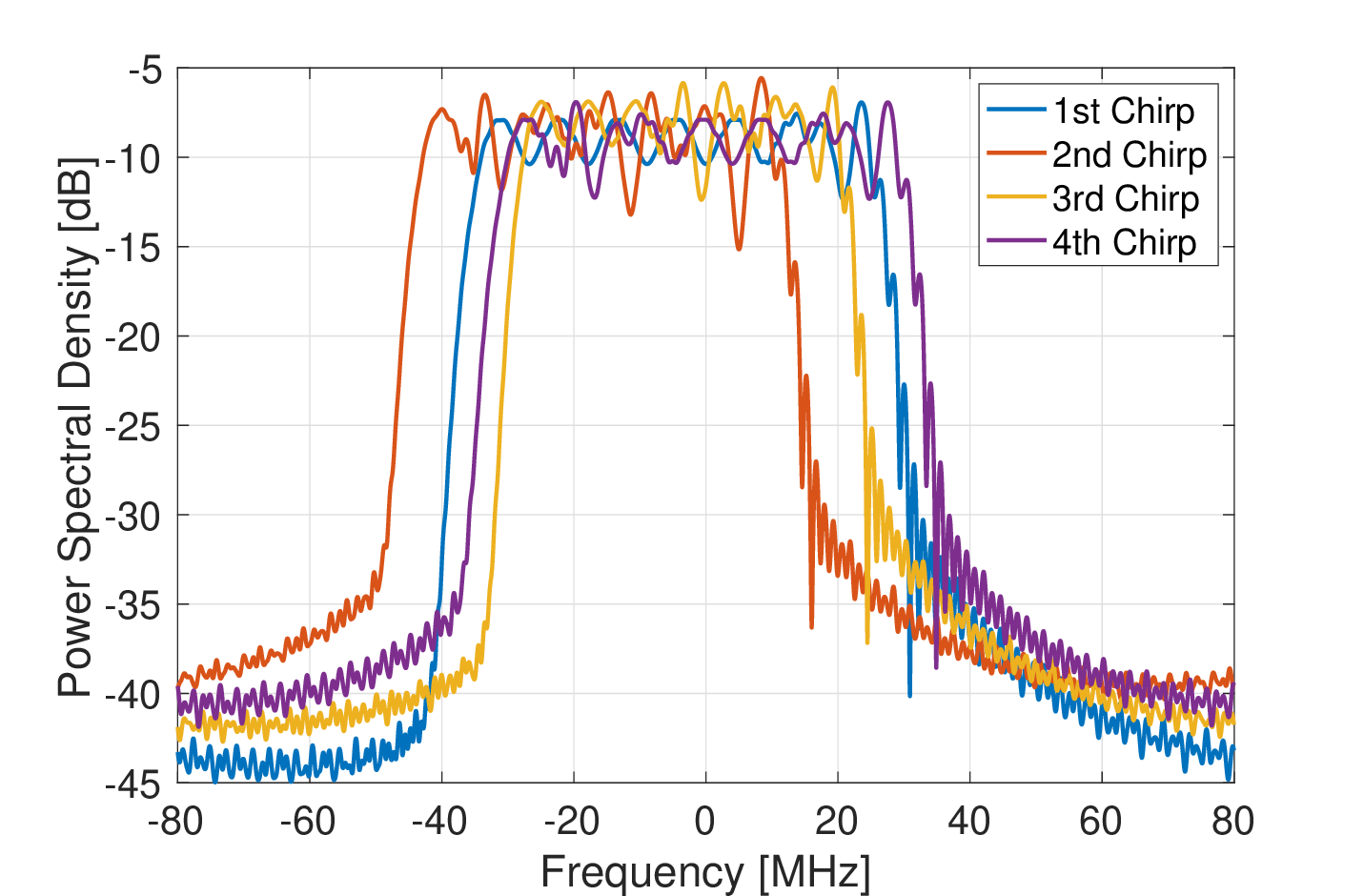} 
        \caption{Frequency Domain}
    \end{subfigure}
    \hfill
    \begin{subfigure}{0.5\textwidth}
        \centering  \includegraphics[width=\textwidth]{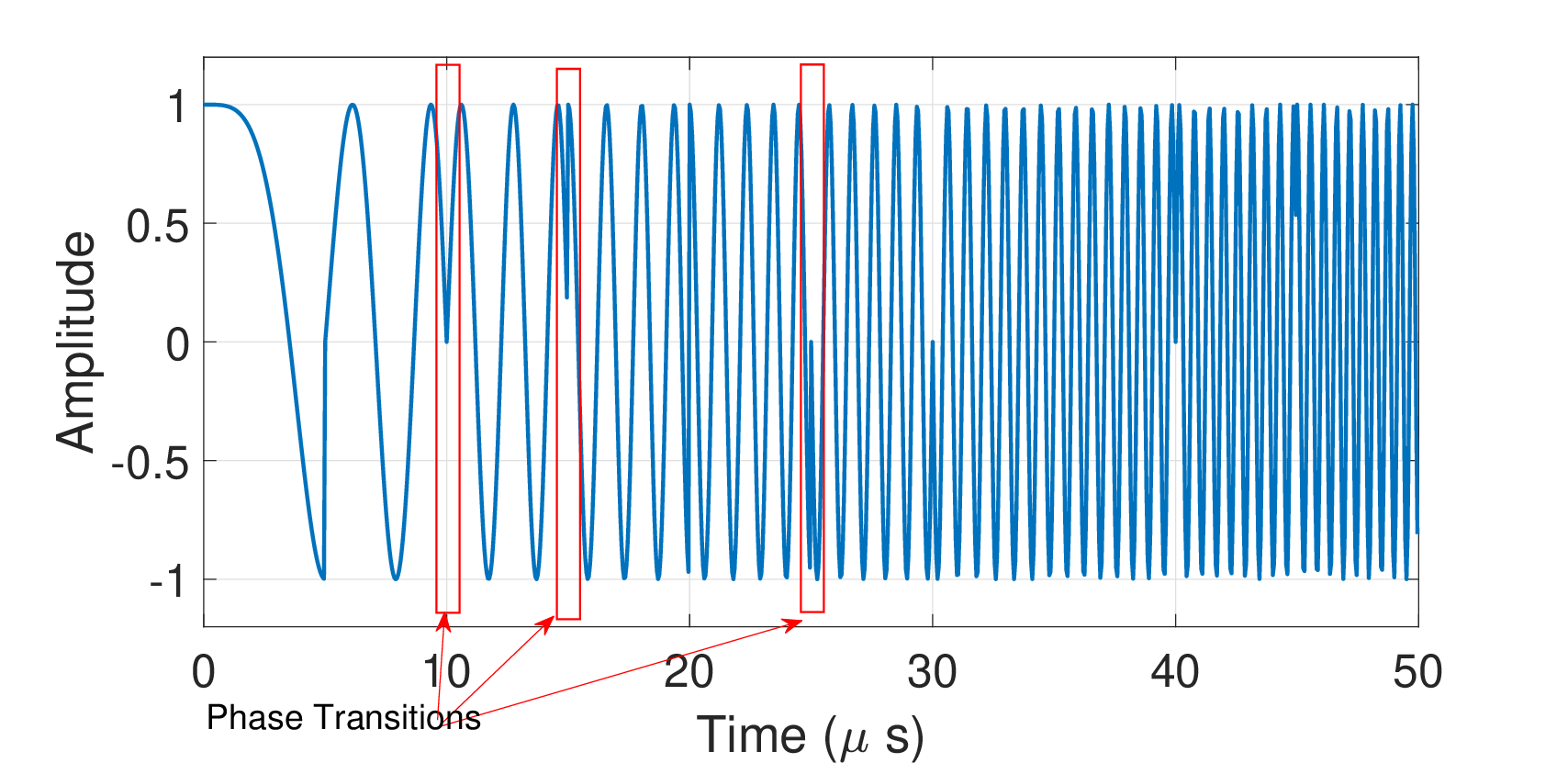} 
        \caption{Time Domain}
    \end{subfigure}
    
    \caption{Illustration of IM-PC-FMCW chirps in the frequency domain and time domain.}
    \label{fig:chirp_freq_time}
\end{figure}

\subsubsection{Index Modulation}

In $x_i(t)$ given by (\ref{fmcw_cis}), the instantaneous frequency $\theta_i(t)$ can be expressed in terms of the chirp bandwidth and the center frequency. Let $b_i$ and $f_i$ denote the bandwidth and center frequency of the $i$th chirp, respectively, where the $[b_i, f_i]$ pair corresponds to the IM indices of the $i$th chirp. In this case, the start and end frequency of the $i$th chirp is given by $f_0^i = f_i-b_i/2$ and $f_1^i = f_i+b_i/2$. Hence, the instantaneous frequency of the chirp, $\theta_i(t)$, is given  by,
\begin{equation}
    \theta_i(t) = \pi\left[\frac{f_1^i-f_0^i}{2}t^2+2f_0^it\right], \quad 0\leq t \leq T,\label{eq_angle}
\end{equation} by replacing $f_i$ and $b_i$ in (\ref{eq_angle}), the following equation is obtained,
\begin{equation}
    \theta_i(t) = \pi\left[\frac{b_i}{2}t^2+2(f_i-\frac{b_i}{2})t\right], \quad 0\leq t \leq T,\label{eq_angle_2}
\end{equation}
where $T_c$ denotes the chirp duration.  Accordingly, the complex-valued $i$th chirp can be generated by \eqref{eq_angle_2} with the desired center frequency $f_i$ and bandwidth $b_i$, corresponding to IM indices.

\subsubsection{Phase Coding}
 The single chirp given by (\ref{fmcw_cis}) is divided into $L$ segments, and the duration of each of the segments is $T/L$. Each segment can have a distinct phase within a chirp. The real and imaginary parts of the phase of the $l$th segment are denoted by $\phi_{re}^l(t)$ and $\phi_{im}^l(t)$, respectively. The phases of the real and imaginary parts of the signal are kept the same for each $l$th segment to avoid unwanted negative frequencies, hence $\phi_{re}^l(t) = \phi_{im}^l(t) = \phi_i^l(t)$. The sequence of phases within a chirp, i.e., phases of $L$ segments, is defined as phase codes. Thus, the transmitted phase code during the $i$th chirp duration is given by
\begin{equation}
    \Phi_i = \left[\phi_i^{1}, \phi_i^{2},\dots, \phi_i^{L}\right],
\end{equation}
where 
\begin{equation}
    \phi_i^l \in  \mathcal{S}_{\mathrm{PSK}} =
    \left\{ \frac{2\pi m}{M} \mid m = 0,1, \dots, M-1 \right\}.\label{eq:PSK_set}
\end{equation} for M-PSK modulation, where the order of the phase modulation in each phase segment of the IM-PC-FMCW signal is denoted by $M$, hence the total data transmitted via phase codes (PC) in each chirp is given by $L \log_2(M)$. By substituting (\ref{eq_angle_2})  in (\ref{fmcw_cis}) and utilizing the Euler's formula, baseband  $i$th IM-PC-FMCW chirp is given by,
\begin{equation}
    x_i(t) = \exp{\left
    (-j \left( \pi\left[\frac{b_i}{2}t^2+2(f_i-\frac{b_i}{2})t\right]+\phi_i^{l}(t)\right)\right)},\label{eq:imp-pc-fmcw}
\end{equation}
where $0\leq t \leq T$ and $l=1, 2, \dots, L$ for each chirp.

Fig.~\ref{fig:chirp_freq_time} illustrates example IM-PC-FMCW chirps in the frequency domain and time domain, where the center frequency and bandwidth of chirps and phase code along a chirp are shown. Accordingly, the data codeword that can be transmitted within a single chirp in V-pol or H-pol can be given as,
\begin{equation}
    \boldsymbol{\Omega}^V_i = \left[f^V_i, b^V_i, \Phi^V_i \right] \quad \text{and} \quad \boldsymbol{\Omega}^H_i = \left[f^H_i, b^H_i, \Phi^H_i \right],
\end{equation}
where each parameter is drawn from a modulation codebook which is designed to enable secure data transmission and improve sensing privacy, as explained below.

\section{Phase Coding for Sensing Privacy}

In ISAC systems, both communication security and sensing privacy must be considered while designing the entire system. In this context, sensing privacy aims to prevent ISAC signals from being used for passive sensing by an adversarial hardware, S-Eve. For passive sensing, S-Eve utilizes a reference signal directly received from the legitimate receiver and also echo signals reflected from the target, as shown in Fig.~\ref{fig:scenario}. The bandwidth and center of the proposed secure IM-PC-FMCW (Sec-FMCW) signals vary from chirp to chirp, and AFs of chirps are specifically engineered to deter the range and velocity estimation abilities of S-Eve. 

\subsection{Ambiguity Function (AF) Design for Sensing Privacy}

The AF characterizes the response of a radar waveform to a time delay \( \tau \) and Doppler frequency shift \( f_D \), and the AF of signal $x$ is given by
\begin{equation}
    \psi_x(\tau, f_D) = \int_{-\infty}^{\infty} x(t) x^*(t - \tau) e^{j 2 \pi f_D t} dt,
\end{equation}
where \( x(t) \), \( \tau \), and \( f_D \) denote the transmitted signal, the time delay (related to the range), and the Doppler shift (related to velocity), respectively. Moreover, \( x^*(t) \) represents the complex conjugate of \( x(t) \). 

The velocity estimation ability of S-Eve is substantially reduced due to arbitrary phase variations that are introduced by varying the center frequency and bandwidth of chirps. Thus, as will be shown in the results, S-Eve cannot estimate the velocity of the target even with high signal-to-noise ratios (SNRs). However, it can still estimate the range of the target, although it is significantly impaired by the varying chirp bandwidth and frequency.  To further reduce S-Eve's range estimation abilities, the phase codes in the signal are designed to cause sidelobes in the range (delay) AF in the proposed Sec-FMCW signaling, as explained below.

The range AF considers only the effect of time delay and is given by,
\begin{equation}
    \psi_x(\tau) = \left|\int_{-\infty}^{\infty} x(t) x^*(t - \tau) dt\right|,
\end{equation}
which determines the ability to resolve targets at different ranges. The range AF, $\boldsymbol{\psi}_x\in \mathcal{R}^{N\times 1}$, can be calculated for the discrete signal vector $x$ with size $N$ as
\begin{equation}
\boldsymbol{\psi}_x[k] = \left|\sum_{n=0}^{N-1} \mathbf{x}[n] \mathbf{x}^*[n-k]\right|, \quad k = 0, 1, \dots, N-1,
\end{equation}
for the $k$th sample in the time domain. However, this calculation is computationally inefficient since it requires an iterative method. The calculation of this can also be performed using the discrete Fourier transform (DFT) and inverse discrete Fourier transform (IDFT) as
\begin{equation}
\boldsymbol{\psi}_x[k] = \left|\frac{1}{N} \sum_{k=0}^{N-1} \left|\sum_{n=0}^{N-1} \mathbf{x}[n] e^{-j 2 \pi kn / N}\right|^2 e^{j 2 \pi km / N}\right|,
\end{equation} and this can be efficiently calculated using fast Fourier transform (FFT) and inverse fast Fourier transform (IFFT) as 
\begin{equation}
    \boldsymbol{\psi}_x =\left| \mathcal{F}^{-1} \left( |\mathcal{F} (\mathbf{x})|^2 \right)\right|,\label{eq:AF_calculation}
\end{equation}
where $\mathcal{F}$ and $\mathcal{F}^{-1}$ denotes the FFT and IFFT operators, respectively.

Let $\boldsymbol{\psi}_d\in \mathbb{R}^{N\times 1}$ denote the desired range AF for sensing privacy that has artificial sidelobes to confuse the S-Eve. To create these sidelobes in the desired location of the range AF, the phase code  $\Phi_i$ within the $i$th Sec-FMCW chirp is optimized as,
\begin{equation}
\begin{aligned}
    \underset{\Phi_i}{\min} \quad &
    \left\|
    \boldsymbol{\psi}_{x,i} - \boldsymbol{\psi}_d
    \right\|_2^2 \\
    \text{s.t.} \quad &
    \boldsymbol{\psi}_{x,i} = \left|\mathcal{F}^{-1}\!\left( \left| \mathcal{F}(\mathbf{x}_i) \right|^2 \right)\right|, \\
    & \phi_i^l \in \mathcal{S}_{\mathrm{PSK}} \quad \forall l.
\end{aligned}
\label{cb_opt}
\end{equation}
where $\Phi_i = \left[\phi_i^{1}, \phi_i^{2},\dots, \phi_i^{L}\right]$, and the set of phase options, $\mathcal{S}_{\mathrm{PSK}}$, is given by \eqref{eq:PSK_set} for phase modulation order $M$.

To create a codebook, the phase codes of all chirps need to be jointly designed. To this end, let  $\boldsymbol{\zeta}$ denote the  codebook consisting of phase codes (codewords) for all chirps to be designed, as
\begin{equation}
    \boldsymbol{\zeta} = \begin{bmatrix}
           \Phi_1 \\
           \Phi_2 \\
           \vdots \\
           \Phi_G
         \end{bmatrix} = \begin{bmatrix}
           \phi_1^{1}, \phi_1^{2},\dots, \phi_1^{L} \\
           \phi_2^{1}, \phi_2^{2},\dots, \phi_2^{L} \\
           \vdots \\
           \phi_G^{1}, \phi_G^{2},\dots, \phi_G^{L}
         \end{bmatrix},    
\end{equation}
where $G$ denotes the number of codewords in the codebook such that $\log_2(G)$ bits can be transmitted in each chirp. The entire codebook can then be designed by



\begin{equation}
\begin{aligned}
\underset{\boldsymbol{\zeta}}{\mathrm{min}} \quad &
    \sum_{i=1}^{G}
    \min_{\boldsymbol{\psi}_{d,k}\in\mathcal{D}}
    \left\|
    \boldsymbol{\psi}_{x,i}
    -
    \boldsymbol{\psi}_{d,k}
    \right\|_2^2  \\
    \text{s.t.} \quad
    & \boldsymbol{\psi}_{x,i}
    =
    \left|
    \mathcal{F}^{-1}\!\left(
    \left|
    \mathcal{F}(\mathbf{x}_i)
    \right|^2
    \right)
    \right|,
    \quad \forall i, \\
    & \min_{\boldsymbol{\psi}_{d,k}\in\mathcal{D}}
    \left\|
    \boldsymbol{\psi}_{x,i}
    -
    \boldsymbol{\psi}_{d,k}
    \right\|_2^2
    \le \varepsilon,
    \quad \forall i, \\
    & \phi_i^l \in \mathcal{S}_{\mathrm{PSK}},
    \quad \forall i,\forall l, \\
    & \|\Phi_i-\Phi_j\|_0 \ge 1,
    \quad \forall\, i\neq j.
\end{aligned}
\label{eq:codebook_opt_hard}
\end{equation}
where $\mathcal{D} =
    \left\{
    \boldsymbol{\psi}_{d,1},
    \boldsymbol{\psi}_{d,2},
    \dots,
    \boldsymbol{\psi}_{d,Z}
    \right\}$ 
denotes the predefined library of $Z$ reference range AFs, where $\boldsymbol{\psi}_{d,k}$ denotes the $k$th reference range AF associated with a desired sidelobe pattern. The optimization problem in \eqref{eq:codebook_opt_hard} jointly designs the entire codebook of $G$ phase-coded chirps. For each codeword, the resulting range AF is matched to a reference AF in the library $\mathcal{D}$, and the total mismatch across all codewords is minimized. The second constraint guarantees that the mismatch of each codeword does not exceed the prescribed deviation level $\varepsilon$, while the last constraint ensures that all codewords in the codebook are distinct. In this way, the designed codebook can realize multiple desired sidelobe patterns while satisfying the PSK alphabet and mismatch constraints. The reference AFs in $\mathcal{D}$ should be designed such that their sidelobe peaks lie at comparable range offsets, so that the resulting waveforms generate strong ghost target effects.

To solve the codebook design problem in \eqref{eq:codebook_opt_hard}, a discrete block coordinate-descent algorithm is proposed. Since each chip phase is constrained to the finite $M$-PSK alphabet, the optimization is carried out by sequentially updating one chip at a time while fixing all remaining chips and all other codewords. Let the $i$th codeword at iteration $\xi$ be denoted by
\begin{equation}
\Phi_i^{(\xi)}=
\left[
\phi_i^{1,(\xi)},\phi_i^{2,(\xi)},\dots,\phi_i^{L,(\xi)}
\right]^{\mathrm T}.
\end{equation}
For the $i$th codeword, the local objective is defined as
\begin{equation}
J_i(\Phi_i)=
\min_{\boldsymbol{\psi}_{d,k}\in\mathcal D}
\left\|
\boldsymbol{\psi}_{x,i}
-
\boldsymbol{\psi}_{d,k}
\right\|_2^2,
\end{equation}
where $\boldsymbol{\psi}_{x,i}$ is obtained from $\Phi_i$ through \eqref{eq:AF_calculation}. At each coordinate-descent step, the $l$th chip of the $i$th codeword is updated as
\begin{equation}
\phi_i^{l,(\xi+1)}
=
\arg\min_{\phi_c \in \mathcal{S}_{\mathrm{PSK}}^{(i,l)}}
J_i\!\left(\widetilde{\Phi}_i^{(\xi,l)}(\phi_c)\right),
\end{equation}
where $\widetilde{\Phi}_i^{(\xi,l)}(\phi_c)$ denotes the candidate codeword obtained by replacing the $l$th chip of $\Phi_i^{(\xi)}$ with $\phi_c$ while leaving all other chips unchanged. The feasible candidate set is given by
\begin{equation}
\begin{aligned}
\mathcal{S}_{\mathrm{PSK}}^{(i,l)}
=
\Bigl\{
\phi_c\in\mathcal{S}_{\mathrm{PSK}}
\;\Big|\;&
\|\widetilde{\Phi}_i^{(\xi,l)}(\phi_c)-\Phi_j\|_0 \ge 1,\ \forall j\neq i, \\
&
J_i\bigl(\widetilde{\Phi}_i^{(\xi,l)}(\phi_c)\bigr)\le \varepsilon
\Bigr\}.
\end{aligned}
\end{equation}
Thus, for each chip position $l$, all $M$ candidate PSK phases are examined, but only those that satisfy both the distinctness constraint and the prescribed mismatch threshold are retained in the feasible set. Among these feasible candidates, the phase yielding the minimum nearest-reference AF mismatch is selected. The updates are repeated over all chips and all codewords until no further reduction in the objective is obtained or a prescribed maximum number of iterations is reached. The worst-case time complexity of the proposed algorithm is dominated by the distinctness checks and scales as $\mathcal{O}\!\left(\Lambda G^2L^2M\right)$, where $\Lambda$ denotes the number of iterations.

\begin{figure}
    \centering
    \includegraphics[width=1\linewidth]{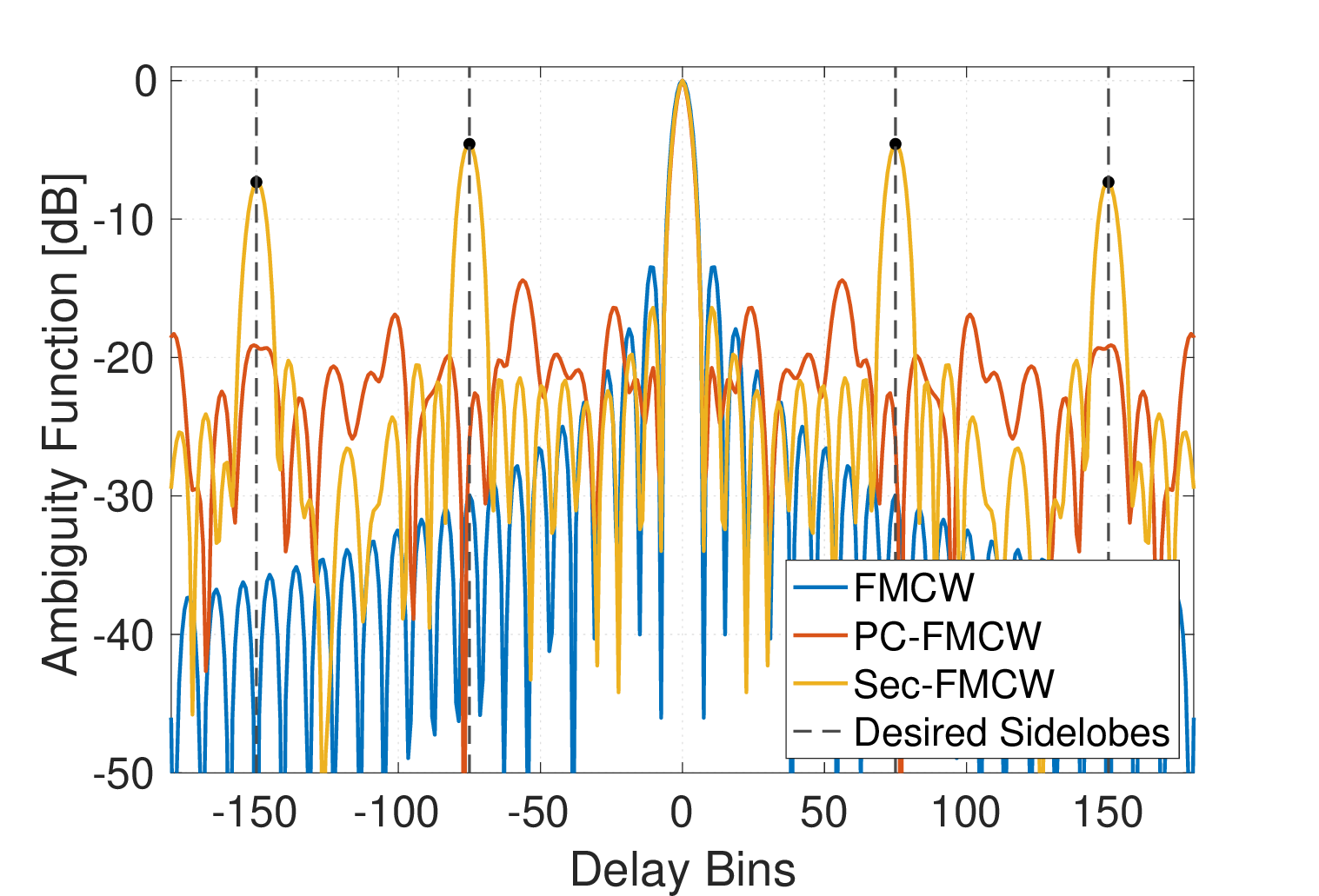}
    \caption{The AFs of IM-FMCW, IM-PC-FMCW and Sec-FMCW  chirps, $T_c=10~\mu$s, $b_i=20$ MHz.}
    \label{fig:optimum_AF}
\end{figure}

Fig.~\ref{fig:optimum_AF} depicts the AFs of the optimized Sec-FMCW signal for sensing privacy in comparison with the FMCW and IM-PC-FMCW signals. As seen in this figure, the Sec-FMCW has sidelobes in the desired locations of the AF, which are much higher than the sidelobes of the random phase-coded IM-PC-FMCW chirp.  In addition to the data transmitted via phase codes, the proposed Sec-FMCW signaling utilizes index modulation, which combines the center frequency and bandwidth selections of each chirp. Moreover, the IM data has to be demodulated first since the demodulation of phase codes requires the knowledge of the center frequency and bandwidth of each chirp. Phase coding and index modulation are combined to create the data codewords. Accordingly, the data codewords during one chirp duration are given by
\begin{equation}
\mathbf{\Omega} = \left\{ \left(f_u, b_u, \Phi_g\right) : u = 1,\ldots,U,\; g = 1,\ldots,G \right\},
\end{equation}
where each codeword is chosen as a combination of $U$ IM indices and $G$ phase codes. Note that the same codebook can be used for both polarizations, hence $\mathbf{\Omega}^V=\mathbf{\Omega}^H=\mathbf{\Omega}$.

\subsection{AF Sidelobe Levels}

To evaluate the sensing performance of the signals, integrated sidelobe level (ISL) and peak sidelobe level (PSL) are used. The ISL quantifies the ratio of the total energy contained within the sidelobes to that of the mainlobe, serving as a key indicator of sensitivity in distributed clutter environments. In contrast, the PSL measures the ratio of the peak sidelobe energy to the mainlobe energy, which determines the system's ability to resolve weak targets in the presence of strong interference. ISL is given by
\begin{equation}
    \text{ISL} = \dfrac{\displaystyle \sum^{N}_{k=-N}\mathbb{E}\left[\left|\psi_x[k]\right|^2\right]-\displaystyle \sum^{k_t}_{k=-k_t}\mathbb{E}\left[\left|\psi_x[k]\right|^2\right]}{\displaystyle \sum^{k_t}_{k=-k_t}\mathbb{E}\left[\left|\psi_x[k]\right|^2\right]},
\end{equation} where $k_t$ denotes the boundary of the mainlobe samples within the AF. The PSL is given by
\begin{equation}
    \text{PSL} = \dfrac{\displaystyle \sum^{k_2}_{k=k_1}\mathbb{E}\left[\left|\psi_x[k]\right|^2\right]}{\displaystyle \sum^{k_t}_{k=-k_t}\mathbb{E}\left[\left|\psi_x[k]\right|^2\right]},
\end{equation} where $k_1$ and $k_2$ denote the lower and upper boundaries of the samples of the highest sidelobe within the AF.

\section{Legitimate Radar Receiver Architecture}\label{sec:legitimate_radar}

The lower part of the ISAC transceiver shown in Fig.~\ref{fig:Isac_transceiever} corresponds to the radar receiver, which receives the echoes from the targets and performs sensing. In a dual-polarized implementation, the received echoes in the V-pol and H-pol branches are down-converted and digitized separately. For notational simplicity, the following signal model is written for a single polarization branch; the same processing applies to the other branch.

Let \(x_i(t)\) denote the complex-baseband signal of the \(i\)-th IM-PC-FMCW, or Sec-FMCW signal given by (\ref{eq:imp-pc-fmcw}). 
Assuming that there are \(K\) targets within the radar field of view, the complex-baseband received signal corresponding to the \(i\)-th chirp can be expressed as
\begin{equation}
    r_i(t)
    =
    \sum_{k=1}^{K}
    \alpha_{T,k}\,
    x_i\!\left(t-\tau_{T,k}\right)
    e^{j2\pi \nu_{T,k} iT_r}
    + n_i(t),
\end{equation}
where \(\tau_{T,k}=2R_{T,k}/c\) is the two-way delay of the \(k\)-th target, \(R_{T,k}\) is its range, \(\nu_{T,k}\) is its Doppler frequency, \(T_r\) is the pulse repetition interval, and \(n_i(t)\sim\mathcal{CN}(0,\sigma_n^2)\) denotes complex additive white Gaussian noise. The complex coefficient \(\alpha_{T,k}\) includes the propagation loss, radar cross section, and the constant carrier-phase term associated with the \(k\)-th target.

For a standard FMCW radar, the beat signal is formed by deramping the received signal with the conjugate of the transmitted chirp, i.e.,    $y_i(t)=r_i(t)x_i^{*}(t).$
However, in the proposed Sec-FMCW scheme, the chirp-dependent bandwidths, center frequencies, and phase codes distort both the fast-time range response and the slow-time phase progression. Therefore, the legitimate radar receiver performs the following processing stages:
\begin{enumerate}
    \item Regularized Inverse Filtering 
    \item Varying Bandwidth Correction,
    \item Center Frequency-hop Phase Correction,
    \item Slow-time FFT for Doppler Estimation.
\end{enumerate}

Accordingly, the proposed radar signal processing is explained as follows.

\subsubsection{Regularized Inverse Filtering}

Let \(r_i[n]\), \(n=0,1,\dots,N_s-1\), denote the sampled received signal of the \(i\)-th chirp, and let \(x_i[n]\) denote the corresponding exact sampled transmitted signal known to the legitimate receiver. The receiver estimates the fast-time response through regularized inverse filtering in the frequency domain. In doing so, let
\begin{equation}
    R_i[k]= \mathcal{F}\left\{r_i[n]\right\}, ~~~ \text{and} ~~~\space X_i[k]= \mathcal{F}\left\{x_i[n]\right\},
\end{equation}
denote the FFTs of the received signal and the exact transmitted signal, respectively. Then, the regularized inverse-filtered spectrum is given by
\begin{equation}
    \hat{Y}_i[k]
    =
    \frac{R_i[k]X_i^{*}[k]}
    {|X_i[k]|^2+\lambda_i},
\end{equation}
where \(\lambda_i>0\) is a regularization parameter that prevents excessive noise enhancement when \(|X_i[k]|^2\) is small. It is selected adaptively as
\begin{equation}
    \lambda_i=\eta\max_k |X_i[k]|^2,
\end{equation}
where \(\eta=0.02\) is used in our implementation. Accordingly, the corresponding fast-time response is obtained by the inverse FFT as
\begin{equation}
    \hat{y}_i[n]
    =
    \mathcal{F}^{-1}\!\left\{\hat{Y}_i[k]\right\}.
\end{equation}
This operation uses the exact waveform knowledge available at the legitimate receiver and mitigates the chirp-dependent spreading caused by the chirp-to-chirp phase coding.

\subsubsection{Varying Bandwidth (Range) Correction}

Let \(N_c\) denote the number of chirps, and let the range domain be discretized into the uniformly spaced estimation vector
\begin{equation}
    \mathbf{r}
    =
    [\,r_0,\;r_1,\;\dots,\;r_{N_r-1}\,]^T,
\end{equation}
where \(N_r\) is the number of range bins. For each hypothesized range bin \(r_m\), the corresponding two-way propagation delay is     $\tau(r_m)={2r_m}/{c}.$  The range response of the \(i\)-th chirp is then written as
\begin{equation}
    \hat{\mathbf{R}}_i
    =
    [\,\hat{R}_i(r_0),\;\hat{R}_i(r_1),\;\dots,\;\hat{R}_i(r_{N_r-1})\,]^T,
\end{equation}
where \(\hat{R}_i(r_m)\) denotes the sampled or interpolated value of \(\hat{y}_i[n]\) at the delay corresponding to \(\tau(r_m)\). Because the inverse filter uses the exact transmitted features of each chirp, the effect of chirp-dependent bandwidth variation is inherently incorporated into the obtained range response.

\subsubsection{Center frequency-hop phase (Doppler) correction}

After inverse filtering and bandwidth compensation, deterministic slow-time phase errors caused by chirp-to-chirp center-frequency and slope variations may still blur the Doppler response. Assume that 
\begin{equation}
    \Delta f_i=f_i-f_{\mathrm{ref}}
\end{equation}
and
\begin{equation}
    \Delta S_i=\frac{b_i-b_{\mathrm{ref}}}{T_c},
\end{equation}
where \(f_{\mathrm{ref}}\) and \(b_{\mathrm{ref}}\) denote the reference center-frequency offset and the reference bandwidth, respectively. Then, the deterministic phase error at range bin \(r_m\) for the \(i\)-th chirp is modeled as
\begin{equation}
    \phi_{\mathrm{err},i}(r_m)
    =
    -2\pi \Delta f_i \tau(r_m)
    + \pi \Delta S_i \tau^2(r_m).
\end{equation}
The corresponding phase-correction factor is given by 
\begin{equation}
    p_i(r_m)=e^{-j\phi_{\mathrm{err},i}(r_m)}.
\end{equation}
Therefore, the phase-corrected range response becomes
\begin{equation}
    \tilde{R}_i(r_m)
    =
    \hat{R}_i(r_m)\,p_i(r_m)
    =
    \hat{R}_i(r_m)e^{-j\phi_{\mathrm{err},i}(r_m)}.
\end{equation}

\subsubsection{Range-Doppler Processing}

After inverse filtering, bandwidth compensation, and residual phase correction, the slow-time sequence at each fixed range bin reflects the true Doppler phase progression. For a given range bin \(r_m\), the compensated responses from all \(N_c\) chirps are
\begin{equation}
    \tilde{R}_0(r_m),\;
    \tilde{R}_1(r_m),\;
    \dots,\;
    \tilde{R}_{N_c-1}(r_m).
\end{equation}
Then, the Doppler spectrum at that range bin is obtained by applying a slow-time FFT across the chirp index:
\begin{equation}
    D(r_m,\ell)
    =
    \sum_{i=0}^{N_c-1}
    \tilde{R}_i(r_m)\,
    e^{-j2\pi \ell i/N_c},
\end{equation}
where \(\ell\) denotes the Doppler-bin index. Repeating this operation for all range bins yields the final focused range-Doppler map.

\subsubsection{Complexity of the Radar Processing}
For the Sec-FMCW signaling, the legitimate radar receiver consists of three main stages: regularized inverse filtering, center-frequency-hop phase correction, and slow-time Doppler FFT. The inverse-filtering stage requires, for each chirp, one FFT of the received signal, one FFT of the exact transmitted reference signals, elementwise regularized inversion, and one IFFT. Therefore, its complexity is approximately \(\mathcal{O}(N_c N_s \log N_s)\), where \(N_c\) is the number of chirps and \(N_s\) is the number of fast-time samples per chirp. After inverse filtering, the carrier-hop phase correction is applied to each range bin of each chirp, which requires \(\mathcal{O}(N_c N_r)\) operations, where \(N_r\) is the number of range bins. Finally, the slow-time Doppler processing performs an \(N_c\)-point FFT for each range bin, leading to a complexity of \(\mathcal{O}(N_r N_c \log N_c)\). Hence, the overall complexity of the proposed legitimate receiver is
\begin{equation}
   \mathcal{O}\!\left(N_c N_s \log N_s + N_c N_r + N_r N_c \log N_c\right).
\end{equation}
For comparison, a conventional matched-filter-based receiver has complexity of $\mathcal{O}\!\left(N_c N_s \log N_s + N_r N_c \log N_c\right)$,
because it does not include the additional chirp-dependent inversion and compensation operations. Although the proposed receiver incurs an extra \(\mathcal{O}(N_c N_r)\) term, this added complexity remains manageable and provides accurate range--Doppler focusing for Sec-FMCW signaling under chirp-dependent waveform variations.

\section{Signal Processing at S-Eve}

Unlike the legitimate radar receiver, S-Eve does not have access to a clean local replica of the transmitted signal or the chirp-dependent parameters required for advanced compensation. Instead, S-Eve is assumed to obtain only a noisy copy of the transmitted signal and therefore relies on conventional matched filtering in the time domain.

Let \(x_i(t)\) denote the transmitted complex-baseband signal of the \(i\)-th chirp, and let the noisy reference received at S-Eve be modeled as
\begin{equation}
x^\mathrm{ref}_{i}(t)
=
\alpha_{\mathrm{ref}}
e^{-j2\pi f_c \tau_{\mathrm{ref}}}
x_i\!\left(t-\tau_{\mathrm{ref}}\right)
+
n_{\mathrm{ref},i}(t),
\end{equation}
where \(\tau_{\mathrm{ref}}\) denotes the propagation delay between the transmitter and S-Eve, \(\alpha_{\mathrm{ref}}\) represents the amplitude attenuation due to the path loss between the ISAC Tx and S-Eve link, \(f_c\) is the carrier frequency, and \(e^{-j2\pi f_c \tau_{\mathrm{ref}}}\) captures the phase shift caused by propagation. Moreover, \(n_{\mathrm{ref},i}(t)\sim\mathcal{CN}(0,\sigma_{\mathrm{ref}}^2)\) denotes additive complex Gaussian noise in the reference channel. The propagation delay can be written as $\tau_{\mathrm{ref}}={d_{\mathrm{ref}}}/{c}$,
where \(d_{\mathrm{ref}}\) is the distance between the transmitter and S-Eve, and \(c\) is the speed of light.

The signal received by S-Eve during the \(i\)-th chirp interval is modeled as
\begin{equation}
r^{\mathrm{eve}}_i(t)
=
\sum_{k=1}^{K}
\alpha_{\mathrm{eve},k}\,
x_i\!\left(t-\tau_{\mathrm{eve},k}\right)
e^{j2\pi \nu_{\mathrm{eve},k} iT_r}
+ n_{\mathrm{eve},i}(t),
\end{equation}
where \(\tau_{\mathrm{eve},k}\) and \(\nu_{\mathrm{eve},k}\) denote the effective delay and Doppler frequency of the \(k\)-th target as observed by S-Eve, \(\alpha_{\mathrm{eve},k}\) is the corresponding complex propagation coefficient, and \(n_{\mathrm{eve},i}(t)\sim\mathcal{CN}(0,\sigma_{\mathrm{eve}}^2)\) is additive complex Gaussian noise.

Since S-Eve does not know the chirp-dependent compensation required by the secure signaling strategy, it performs only time-domain matched filtering between the received signal and the noisy reference signal. Accordingly, the delay-domain response for the \(i\)-th chirp is written as
\begin{equation}
z_i(\tau)
=
\int r^{\mathrm{eve},i}(t)\,
(x^{\mathrm{ref}}_i(t-\tau))^*\,dt,
\end{equation}
where \(\tau\) denotes the delay variable. In discrete time, this operation corresponds to the cross-correlation between the received signal and the noisy reference signal.

After obtaining the delay-domain response for each chirp, S-Eve performs Doppler processing across the slow-time dimension. For a fixed delay bin \(\tau_m\), the Doppler spectrum is obtained as
\begin{equation}
D_{\mathrm{eve}}(\tau_m,\ell)
=
\sum_{i=0}^{N_c-1}
z_i(\tau_m)\,
e^{-j2\pi \ell i/N_c},
\end{equation}
where \(\ell\) denotes the Doppler-bin index. Repeating this operation for all delay bins yields the range-Doppler map observed by S-Eve.

Because S-Eve relies only on a noisy intercepted reference and does not perform regularized inverse filtering or chirp-dependent carrier-hop compensation, its range-Doppler focusing performance is degraded compared with that of the legitimate radar receiver.

\section{Communication Receiver Architecture}

\begin{figure*}
\centering\includegraphics[width=0.8\linewidth]{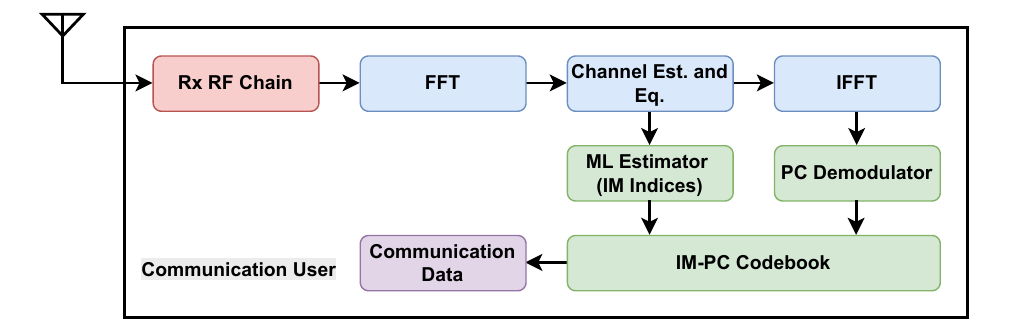}
    \caption{The proposed communication receiver architecture to receive the communication data.}
    \label{fig:Com_receiver}
\end{figure*}

Fig.~\ref{fig:Com_receiver} shows the stages of the demodulation of the communication data performed at the CU, where the RF signals are downconverted to baseband signals and then channel equalization is performed. The legitimate receiver can perform channel estimation by utilizing phase-coded Sec-FMCW pilot chirps and perform equalization before demodulation. However, C-Eve cannot perform channel estimation since it does not know the pseudorandom phase-coded pilot chirps that are arbitrarily generated and only known by the ISAC BS and the legitimate receiver.

\subsection{Channel Estimation}
Both the CU and C-Eve perform channel estimation prior to demodulating the received chirps. As shown in Fig.~\ref{fig:transmission_frame}, pilot Sec-FMCW chirps are inserted periodically within the transmission frame for this purpose. Because each pilot chirp occupies the full allocated bandwidth, a single pilot chirp is sufficient for channel estimation as long as the channel remains approximately constant over the following data-chirp interval. The pilot chirps are encoded with pseudo-random phase codes known to the BS and legitimate users, but unknown to C-Eve. Consequently, the CU can estimate the channel using the exact pilot waveform, whereas C-Eve relies on a standard FMCW signal for channel estimation.

After the pilot chirp, multiple Sec-FMCW data chirps are transmitted for joint secure communication and sensing. The CU uses the estimated channel state information (CSI) to equalize the subsequent data chirps. In addition, radar sensing remains uninterrupted during channel estimation, since both pilot and data chirps can be exploited for sensing. The received frequency-domain pilot signals in the V-pol and H-pol are given by
\begin{align}
\mathbf{y}_p^V &= \mathbf{h}^V\odot\mathbf{u}_p^V+\mathbf{h}^{HV}\odot\mathbf{u}_p^H+\mathbf{n}^V,\notag \\
\mathbf{y}_p^H &= \mathbf{h}^H\odot\mathbf{u}_p^H+\mathbf{h}^{VH}\odot\mathbf{u}_p^V+\mathbf{n}^H,
\end{align}
respectively, where $\mathbf{u}_p^V=\mathcal{FFT}(\mathbf{x}_p^V)$ and $\mathbf{u}_p^H=\mathcal{FFT}(\mathbf{x}_p^H)$ denote the pilot FMCW chirps represented in the frequency domain. The symbols $\odot$ and $\oslash$ indicate element-wise multiplication and division, respectively. The bandwidth of the pilot chirps covers the complete allocated bandwidth, allowing for an estimation of the entire channel response for Sec-FMCW chirps.

{%
The frequency-domain linear minimum mean square error (LMMSE) channel estimator is utilized for channel estimation. Let \(\mathbf{I}_1 \in \mathbb{R}^{N_s \times 1}\) represent an all-ones vector, where \(N_s\) denotes the number of samples within a chirp. Assuming that the noise and polarization statistics are the same for both polarizations, the thermal noise variance and cross-polarization interference variance are represented by \(\sigma_n^2\) and \(\sigma_i^2\), respectively, for both polarizations. The channels in the vertical (V-pol) and horizontal (H-pol) polarizations are estimated as follows,
\begin{align}
\mathbf{h}_e^V &=
\left(\mathbf{y}_p^V\odot(\mathbf{u}_p^V)^*\right)\oslash
\left(\mathbf{u}_p^V\odot(\mathbf{u}_p^V)^*+(\sigma_n^2+\sigma_i^2)\mathbf{I}_1\right), \notag\\
\mathbf{h}_e^H &=
\left(\mathbf{y}_p^H\odot(\mathbf{u}_p^H)^*\right)\oslash
\left(\mathbf{u}_p^H\odot(\mathbf{u}_p^H)^*+(\sigma_n^2+\sigma_i^2)\mathbf{I}_1\right),
\label{eq:lmmse_ce}
\end{align}
where $(\cdot)^*$ denotes complex conjugation.}

{%
This estimated CSI is utilized to equalize subsequent Sec-FMCW chirps until a new pilot FMCW chirp is received, and the channel estimate is updated. Let $\mathbf{y}_i^V$ and $\mathbf{y}_i^H$ denote the received frequency-domain signals for the $i$th Sec-FMCW data chirp. The data chirps are equalized using a minimum mean square error (MMSE) equalizer, which is applied element-wise as
\begin{align}
\hat{\mathbf{u}}_i^V &=
\left((\mathbf{h}_e^V)^*\odot\mathbf{y}_i^V\right)\oslash
\left(\mathbf{h}_e^V\odot(\mathbf{h}_e^V)^*+(\sigma_n^2+\sigma_i^2)\mathbf{I}_1\right), \notag\\
\hat{\mathbf{u}}_i^H &=
\left((\mathbf{h}_e^H)^*\odot\mathbf{y}_i^H\right)\oslash
\left(\mathbf{h}_e^H\odot(\mathbf{h}_e^H)^*+(\sigma_n^2+\sigma_i^2)\mathbf{I}_1\right).
\label{eq:mmse_eq}
\end{align}
Finally, the time-domain equalized chirps are obtained as $\hat{\mathbf{x}}_i^V=\mathcal{IFFT}(\hat{\mathbf{u}}_i^V)$ and $\hat{\mathbf{x}}_i^H=\mathcal{IFFT}(\hat{\mathbf{u}}_i^H)$.}

\subsection{Demodulation of Index Modulation}

After channel estimation and equalization, the communication receiver estimates bandwidth ($b_i$) and center frequency indices ($f_i$) for each chirp received, using a maximum likelihood  (ML) estimator and IM codebook, using two parallel maximum likelihood (ML) estimators, one for each dual-polarized antenna channel (V-pol and H-pol). The implementation of the ML estimators used for IM demodulation is illustrated by Fig.~\ref{fig:Com_receiver}, which is performed in the frequency domain. Let $\mathcal{F}_{IM}$ denote the IM codebook consisting of all possible center frequencies and bandwidths as
\begin{equation}
\mathcal{F}_{IM} = \{\mathbf{u}_{(1,1)}, \dots \mathbf{u}_{(b,f)}, \dots, \mathbf{u}_{(N_{BW},N_{Fc})}\}, 
\end{equation}
where $N_{BW}$ and $N_{Fc}$ denote the number of bandwidth and center frequency options. Using this IM codebook, two ML estimators can concurrently operate to detect the symbols transmitted in V-pol and H-pol as,
\begin{argmaxi}
    {(\hat{b}^V_i,\hat{f}^V_i)}{|\mathbf{u}^V_{i} \mathbf{u}^*_{(b,f)} |^2}
    {}{}
    ,\label{MLestimator2a2}\notag
\end{argmaxi}
\begin{argmaxi}
    {(\hat{b}^H_i,\hat{f}^H_i)}{|\mathbf{u}^H_{i} \mathbf{u}^*_{(b,f)} |^2}
    {}{}
    ,\label{MLestimator2b2}
\end{argmaxi}
where $*$ denotes the complex conjugate, and the ML estimators return the estimated indices $(\hat{b}^V_i,\hat{f}^V_i)$ and $(\hat{b}^H_i,\hat{f}^H_i)$ for the $i$th chirps received in V-pol and H-pol.  


\subsection{Demodulation of Phase Coding}\label{Sec:Demod_PC}

Before demodulating the phase codes, it is necessary to correctly demodulate the IM so that the communication receiver can know the center frequency and bandwidth of the chirps. After taking the IFFT of the frequency-domain chirps, the channel-equalized time-domain signal vector of the $i$th chirp $\hat{\mathbf{x}}_i$ is obtained. After that, reference signals, $\mathbf{x}_{i,p}$ with the same center frequency and bandwidth but different phases $p$ are generated and multiplied by the conjugate of the received chirp. The receiver integrates the multiplication results for each phase segment, $l=1, 2, \dots, L$, of the chirp and estimates the phases within a chirp for V-pol and H-pol as, 
\begin{argmaxi}
    {{\hat{\phi}_l}^V}{|\mathbf{x}^V_{i,l} \mathbf{x}^{V*}_{i,p,l}|^2}
    {}{}
    ,\label{MLestimator2a2Phase}\notag
\end{argmaxi}
\begin{argmaxi}
    {\hat{\phi}_l^H}{|\mathbf{x}^H_{i,l} \mathbf{x}^{H*}_{i,p,l} |^2}
    {}{}
    ,\label{MLestimator2b2Phase}
\end{argmaxi}
where vectors $\mathbf{x}^H_{i,l}$ and $\mathbf{x}^{H}_{i,l}$ denote the $l$th section of vectors $\mathbf{x}^H_{i}$ and $\mathbf{x}^{H}_{i,p}$ for $l=1,2,\dots, L$. Estimated phase vectors are then obtained as $\hat{\Phi}^V_i  = [\hat{\phi}_1^V, \hat{\phi}_2^V,\dots, \hat{\phi}_L^V]$ and $\hat{\Phi}^H = [\hat{\phi}_1^H, \hat{\phi}_2^H,\dots, \hat{\phi}_L^H]$ for V-pol and H-pol channels. After completing demodulation of IM and PC, the final data is obtained by getting the corresponding data from the codebook for the estimated IM and PC data ($\hat{b}^V_i,\hat{f}^V_i, \hat{\Phi}^V$) for V-pol and ($\hat{b}^H_i,\hat{f}^H_i, \hat{\Phi}^H$) for H-pol. After completing demodulating IM and PC, the final data is obtained by drawing the corresponding data for the combination of IM and PC from the codebook $\boldsymbol{\Omega}$.

On the other hand, C-Eve cannot successfully demodulate the chirps, as it is unable to perform channel estimation without knowledge of the arbitrarily encoded pilot chirps. In addition, C-Eve needs to know the chirp bandwidths, center frequencies, and the employed two-layer modulation scheme. Moreover, the codebook that jointly combines index modulation (IM) and phase coding (PC) for data encoding is also unknown to C-Eve. Owing to these limitations, the proposed signaling provides a high level of physical-layer security against eavesdropping.

\section{Communication Security}

The physical layer communication security is provided by a combination of encoded pilot chirps and a codebook consisting of both IM and PC codewords. The pilot chirps are also encoded via IM and PC to reduce the channel estimation accuracy of C-Eve.

\subsection{Maximum Throughput for Sec-FMCW}


Let us assume that the codebook of phase-coded chirps is designed using an $M$-PSK phase alphabet, code length $L$, and a library of $Z$ reference range AFs under a relaxed AF design constraint with deviation factor $\varepsilon$. In this case, each chip phase can take one of the values in
$\mathcal{S}_{\mathrm{PSK}}$ given by \eqref{eq:PSK_set}, whose spacing is $2\pi/M$. Under the deviation constraint induced by $\varepsilon$, each chip phase of a feasible codeword associated with a given reference AF is allowed to deviate from its nominal value by at most $
\Delta_{\varepsilon}=\arcsin(4\varepsilon)$.
Hence, the feasible phase values for one chip must lie within an interval of width $2\Delta_{\varepsilon}$. Since adjacent $M$-PSK phases are separated by $2\pi/M$, the number of admissible PSK phases per chip is upper bounded by
\begin{equation}
    \left\lceil
\frac{2\Delta_{\varepsilon}}{2\pi/M}
\right\rceil + 1
=
\left\lceil
\frac{M}{\pi}\arcsin(4\varepsilon)
\right\rceil + 1.
\end{equation}
Therefore, for a codeword of length $L$, the maximum number of feasible phase-code vectors associated with one reference AF can be expressed as
\begin{equation}
N_{\phi}=
\left(
\left\lceil
\frac{M}{\pi}\arcsin(4\varepsilon)
\right\rceil + 1
\right)^L.    
\end{equation}
Since there are $Z$ reference AFs in the library, the codebook size $G$ is upper bounded by $ZN_{\phi}$, hence,
\begin{equation}
G \le
Z\left(
\left\lceil
\frac{M}{\pi}\arcsin(4\varepsilon)
\right\rceil + 1
\right)^L, \label{eq:codebook_size}   
\end{equation}
where $\lceil . \rceil $ denotes the ceiling function. Without the AF-constrained secure phase-code design, i.e., for IM-PC-FMCW, the codebook size is limited by $M^L$. 

Each codeword $\boldsymbol{\Omega}_i$ (transmitted in V-pol or H-pol) of the codebook $\boldsymbol{\Omega}$ can be constructed by choosing one index from the $U$ IM options (combination of center frequencies and bandwidths), and $G$ phase codewords. Thus, the number of codewords in the codebook $\boldsymbol{\Omega}$  is given by $S = U\times G$. In this case, the maximum throughput that can be achieved by utilizing the chirps transmitted in both V-pol and H-pol is given by
\begin{equation}
    T_{max} = \frac{\lfloor2\log_2 S\rfloor}{T_c}, \label{eq:Rmax}
\end{equation}
where $\lfloor.\rfloor$ defines the floor function and $T_c$ denotes the chirp duration, which is the inverse of chirp repetition frequency, as $T_c=1/F_r$. By using \eqref{eq:codebook_size}, the maximum throughput of Sec-FMCW signaling is given by 
\begin{equation}
T_{\max}^{\text{Sec-FMCW}}
\leq
\frac{
\left\lfloor
2\left[
\log_2(UZ)
+
L\log_2\!\left(
\left\lceil
\frac{M}{\pi}\arcsin(4\varepsilon)
\right\rceil + 1
\right)
\right]
\right\rfloor
}{T_c},
\label{eq:Rmax}
\end{equation} while the maximum throughput of the IM-PC-FMCW signaling is given by
\begin{equation}
T_{\max}^{\text{IM-PC-FMCW}}
=
\frac{\left\lfloor 2\left(\log_2 U + L\log_2 M\right)\right\rfloor}{T_c}.
\label{eq:Rmax_impc}
\end{equation}


\subsection{Throughput Gap Between CU and C-Eve}

The Sec-FMCW chirps transmitted in V-pol and H-pol during a chirp duration need to be correctly demodulated for a correct decoding of the communication data. Moreover, the bit error rates for demodulating V-pol and H-pol signals are identical due to having similar noise and channel gain statistics. Hence, the average throughput, $T$, is  given by, 
\begin{equation}
    T = (1-p_e)T_{max}=\frac{(1-p_e)\lfloor 2 \log_2 S\rfloor}{T_c}, \label{ins_rate}
\end{equation}
where $p_e$ denotes the block error rate of the demodulation using the codebook $\boldsymbol{\Omega}$. It is worth noting that equation (\ref{ins_rate}) states that having a shorter chirp duration, $T_c$, increases the throughput.

The difference between average throughputs $T_u$ and $T_{eve}$, which can be achieved by the CU and C-Eve, is defined as the throughput gap ($T_s$), corresponding to the communication security metric. A higher throughput gap indicates a higher communication security, and the best case can be achieved when $T_{eve}=0$. The throughput gap, ($T_s$), is given by
\begin{align}
    T_s &= T_{u} - T_{eve} \notag \\
    &= \frac{(1-p_e^u)\lfloor 2 \log_2 S\rfloor}{T_c} - \frac{(1-p_e^{eve})\lfloor 2 \log_2 S\rfloor}{T_c}\notag \\
    &=
    \left(p_e^{eve} - p_e^{u}
    \right)
    \frac{\lfloor 2 \log_2 S\rfloor}{T_c}, 
\end{align}
where $0\leq p_e^u\leq1$ and $0\leq p_e^{eve}\leq1$ denote the block error rate for the demodulations performed by the CU and C-Eve, respectively. This shows that the throughput gap can be maximized by minimizing the block error rate of the CU while maximizing the block error rate of C-Eve. 

The maximum throughput gap will be $T_{s}=T_{max}$ as bounded by \eqref{eq:Rmax} when $p_e^u=0$ and $p_e^{eve}=1$, hence $T_{u}=T_{max}$, $T_{eve}=0$. The proposed two-layer modulation scheme and codebook design maximize the codeword estimation errors of C-Eve, hence providing a highly secure physical layer for communications. Moreover, Pilot chirps, which are utilized for channel estimation, are encoded using the pilot codewords to further improve the security of the communications, as shown in Fig.~\ref{fig:transmission_frame}, where phase-coded pilot Sec-FMCW chirps are transmitted for channel estimation after a certain number of data chirps.

\section{Simulations and Results}

\begin{figure}
    \centering \includegraphics[width=0.95\linewidth]{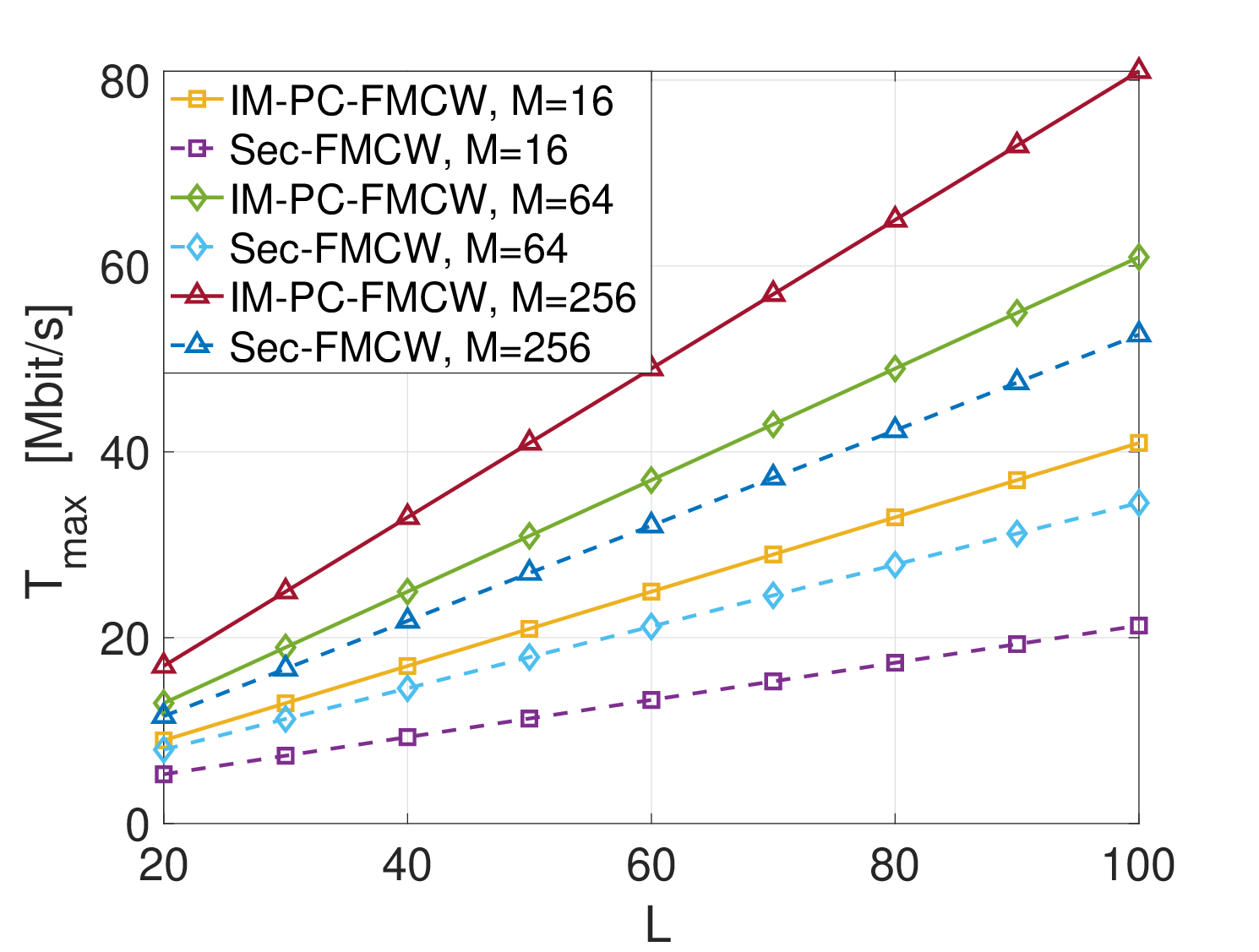}
    \caption{Maximum throughput that can be achieved by CU with IM-PC-FMCW and Sec-FMCW, $T_c=20$ $\mu$s, $Z=10$, $\varepsilon=0.1$.}
    \label{fig:max_throughput}
\end{figure}

\begin{figure}
    \centering \includegraphics[width=0.88\linewidth]{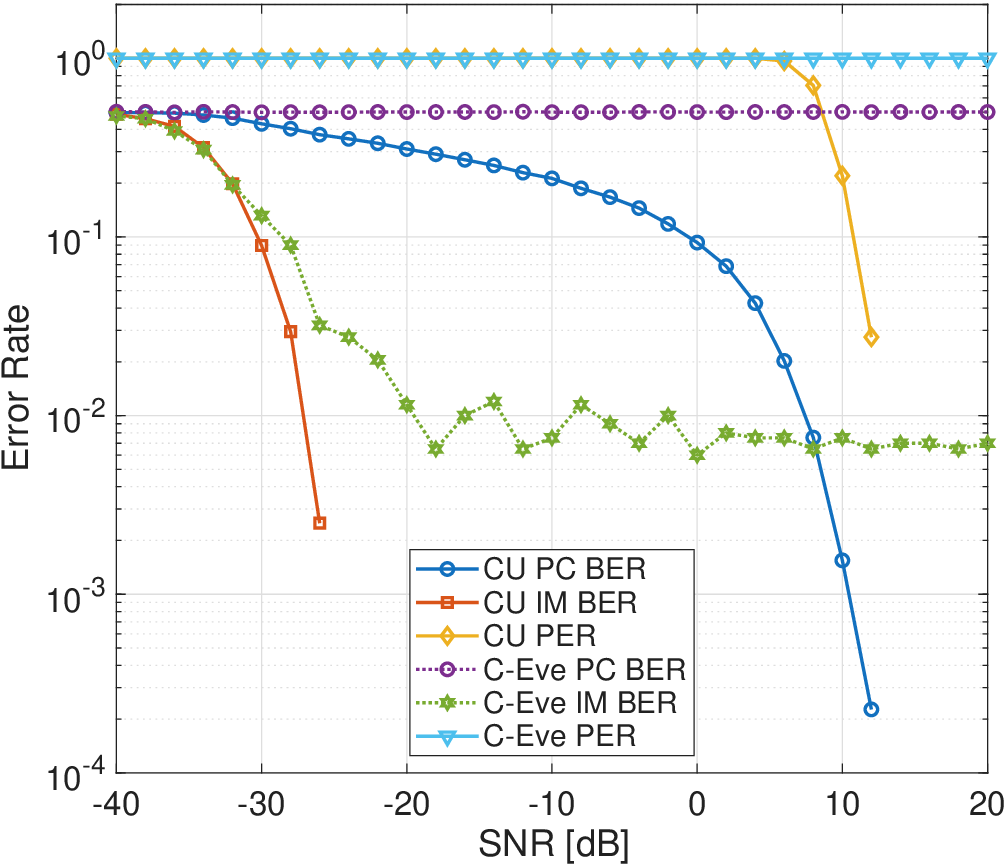}
    \caption{BER and PER of the CU and C-Eve ($M=256$, $L=40$, $T = 20\mu$s, $\Delta F_c = \Delta Bw = 1$MHz).}
    \label{fig:secure_error}
\end{figure}


\begin{figure}
    \centering \includegraphics[width=1\linewidth]{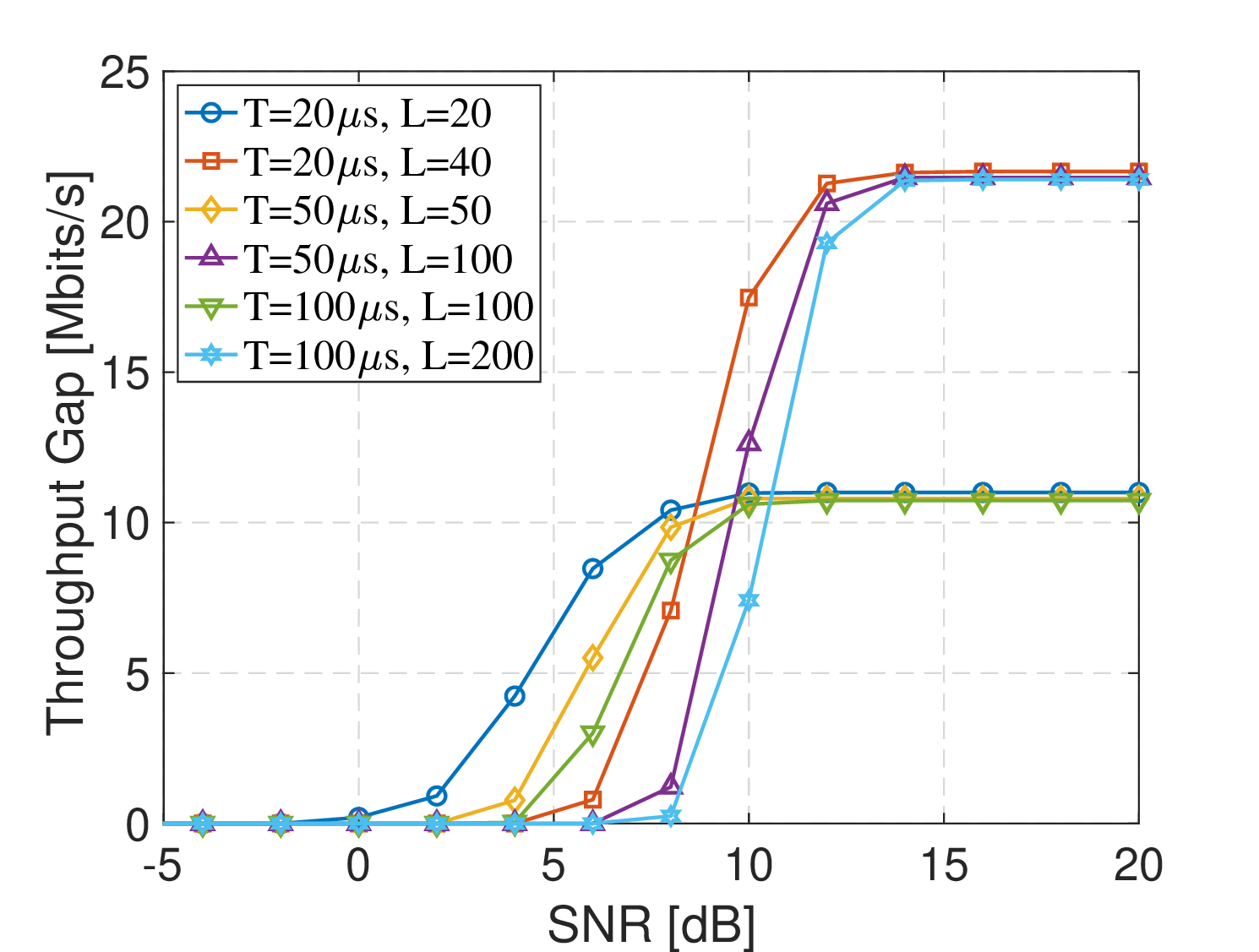}
\caption{Throughput gap of Sec-FMCW signals with various phase code lengths per chirp (M=256).}
\label{fig:secure_rate}
\end{figure}

\begin{figure}
    \centering \includegraphics[width=0.85\linewidth]{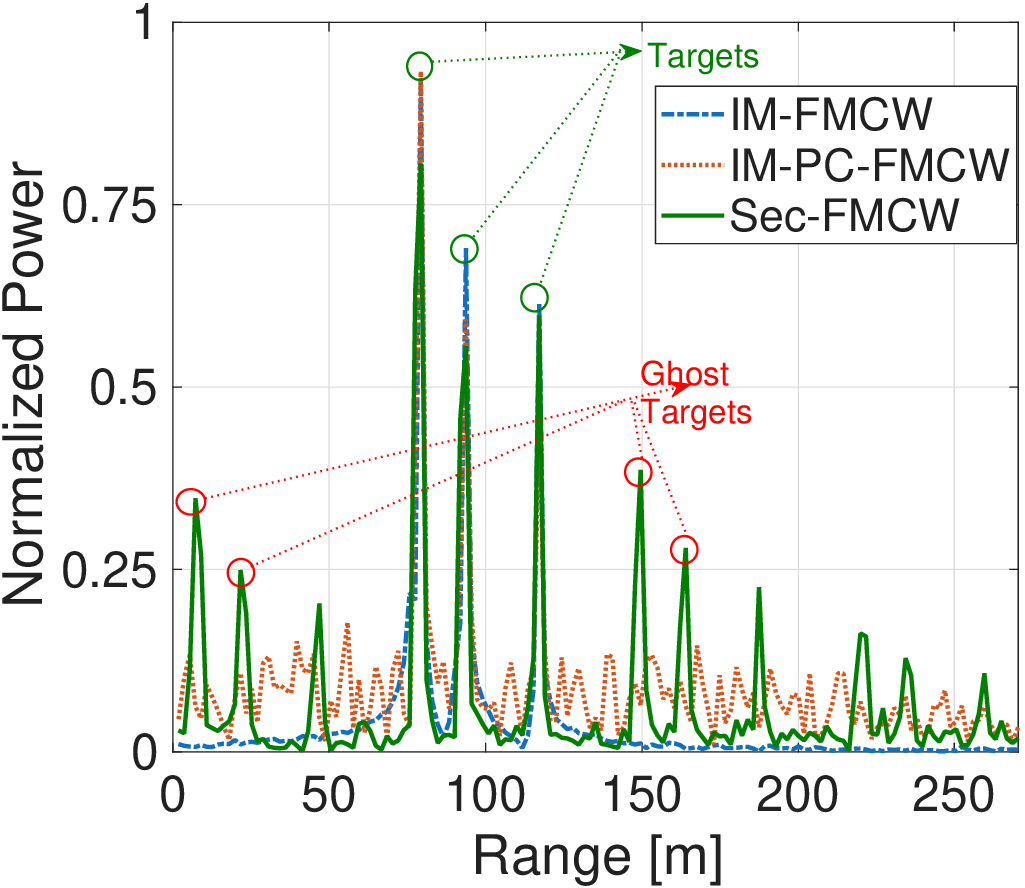}
\caption{Uncompensated matched-filter range profiles obtained at the legitimate receiver using different signals, SNR = 20 dB.}
\label{fig:range_profile}
\end{figure}

\begin{figure}[ht]
    \centering \includegraphics[width=0.92\linewidth]{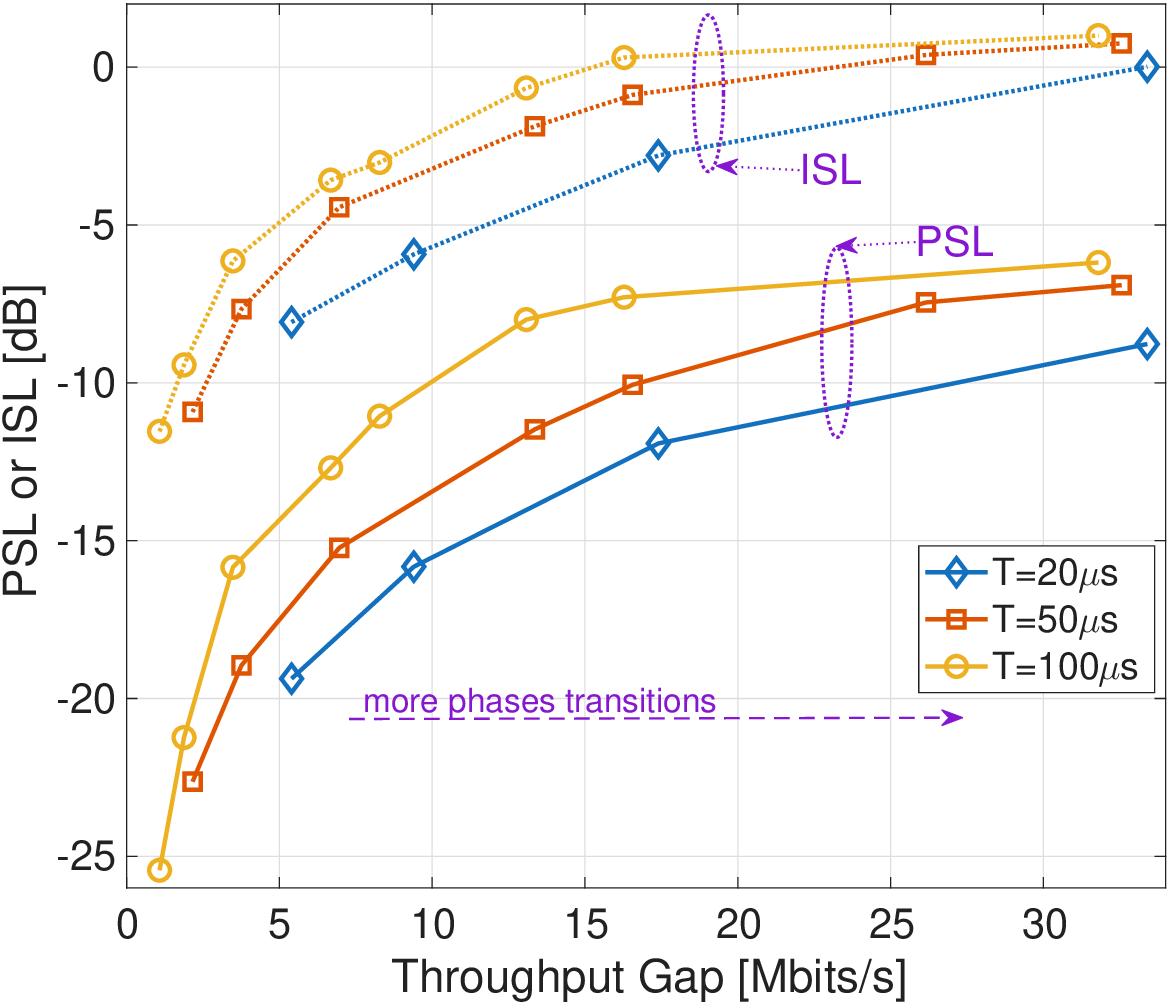}
    \caption{The PSL and ISL versus throughput in relation to the number of phase transition points in each chirp, M=256, $T_c=20$ $\mu$s.}
    \label{fig:trade-off}
\end{figure}

\begin{figure*}
    \centering
\includegraphics[width=0.95\linewidth]{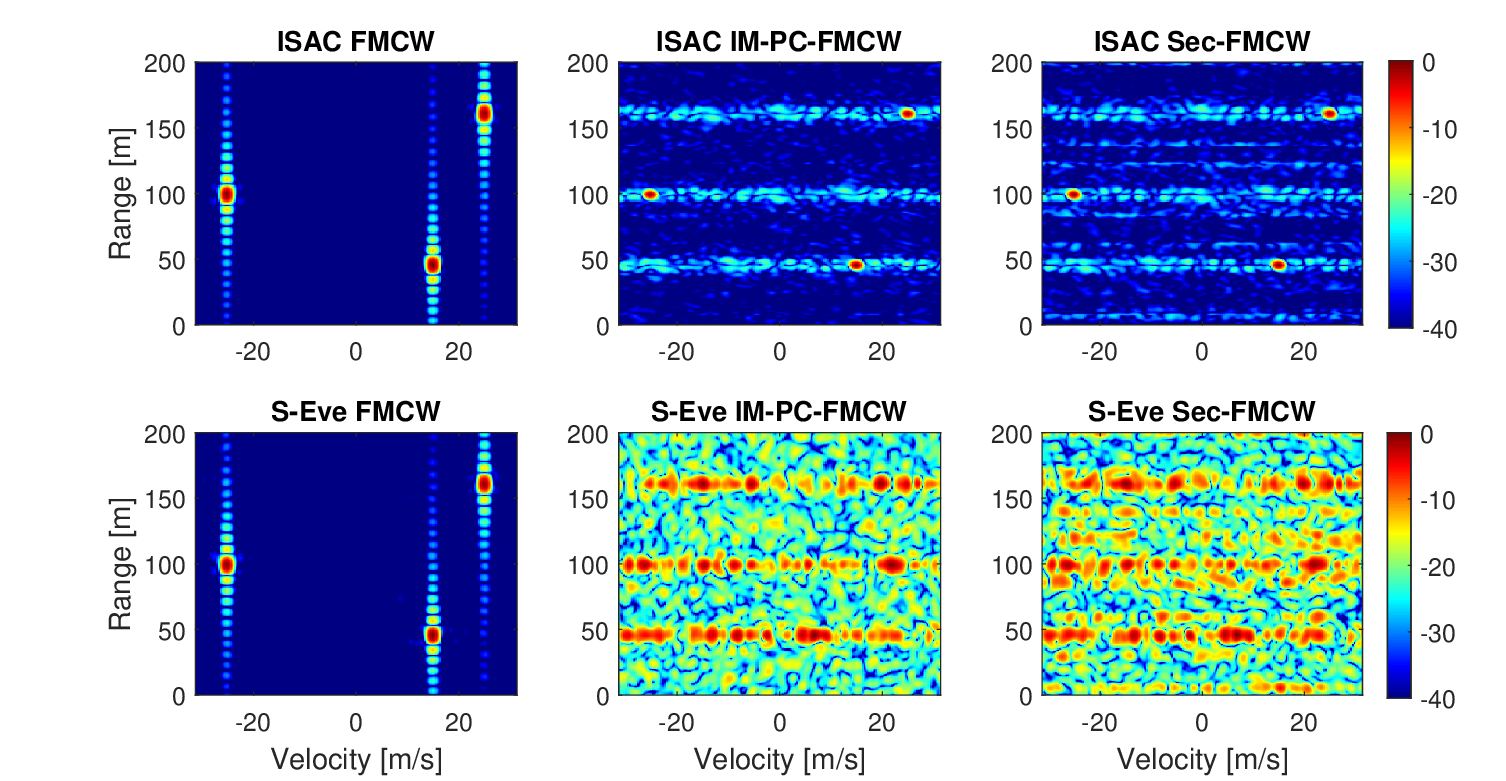}
    \caption{Range-velocity maps obtained by ISAC sensing receiver and S-Eve, SNR = 20 dB.}
    \label{fig:range-velocity-eve}
\end{figure*}

\begin{figure}[h]
    \centering
    \begin{subfigure}{0.47\textwidth}
        \centering  \includegraphics[width=\textwidth]{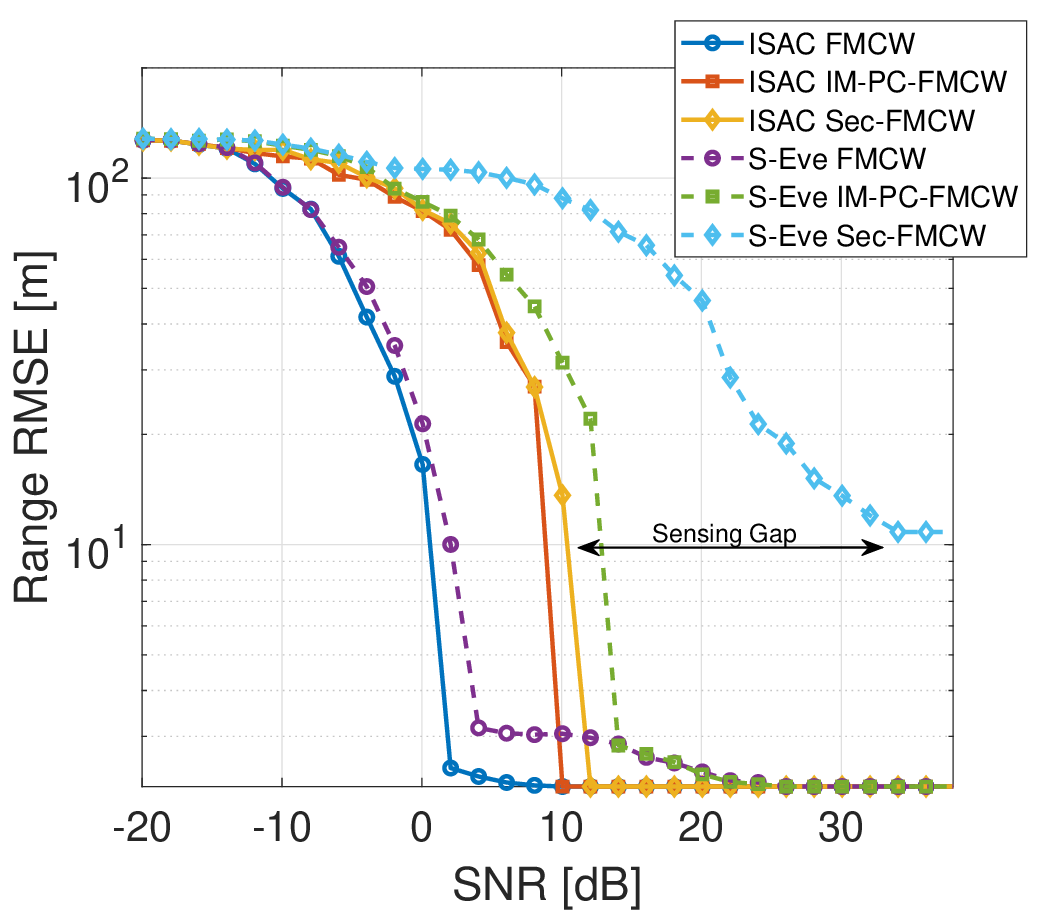} 
        \caption{RMSE of range estimation.}
    \end{subfigure}
    \hfill
    \begin{subfigure}{0.47\textwidth}
        \centering  \includegraphics[width=\textwidth]{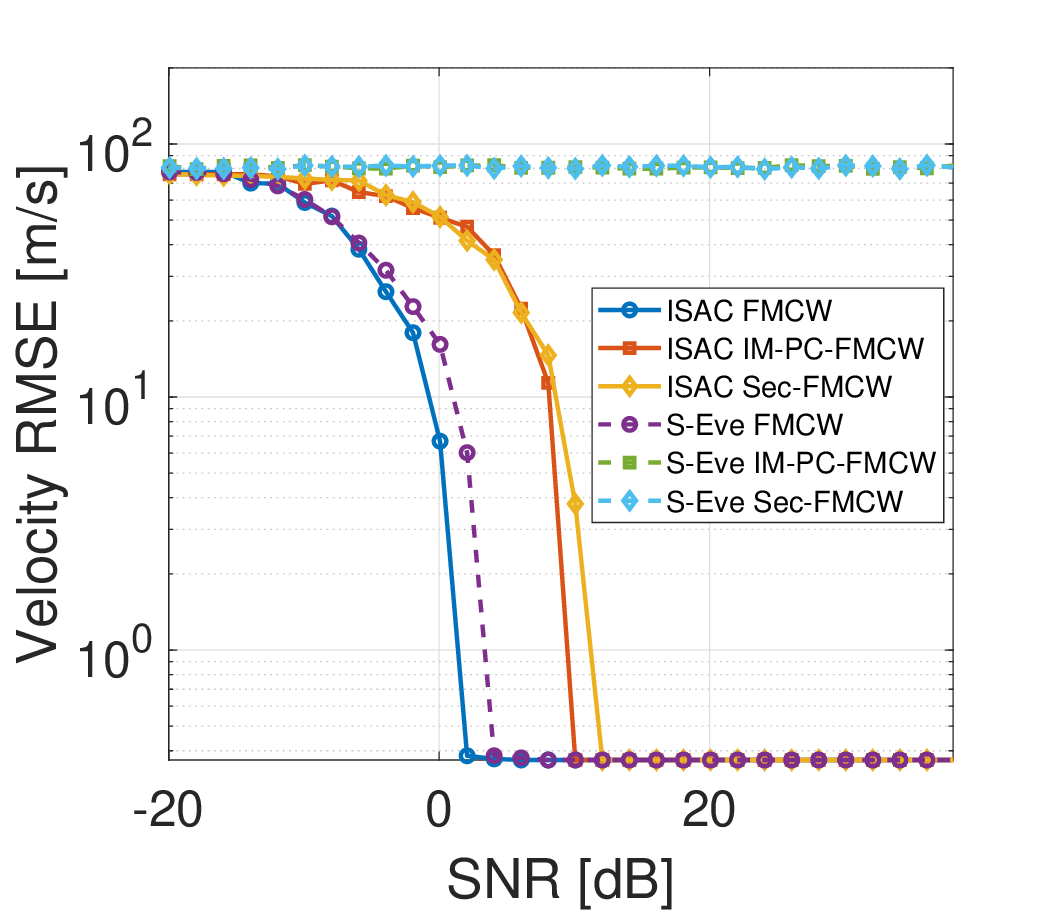} 
        \caption{RMSE of velocity estimation.}
    \end{subfigure}
    \caption{Mean range and velocity errors of the target parameter estimations of ISAC BS and S-Eve.}
    \label{fig:error_stats}
\end{figure}

The simulations are performed to verify the sensing privacy and communication security of the proposed secure signaling. S-Eve is considered to be near the BS, hence it also performs monostatic sensing for a fair comparison. In the simulation setup, a radar system with a 2.4 GHz carrier frequency is considered, where targets with 54 km/h (15 m/s), -90 km/h (-25 m/s), and 90 km/h (25 m/s) velocities within 45 m, 100 m, and 160 m ranges, respectively,  are placed in front of the ISAC BS. 

The bandwidths of the chirps are within 30 MHz and 50 MHz with a minimum 1 MHz separation between the center frequency and bandwidth options of the chirps, i.e., $\Delta f=1$ MHz and $\Delta b=1$ MHz. This results in $U=861$ selections for IM indices. The larger communication codebook $\boldsymbol{\Omega}$ is formed as the combination of the IM and phase codebook, and then arbitrary data is assigned to each codeword in $\boldsymbol{\Omega}$. Moreover, $\varepsilon=0.1$, $Z=10$ are chosen to generate the secure phase codebook for Sec-FMCW. 

Fig.~\ref{fig:max_throughput} shows the maximum achievable throughput of the IM-PC-FMCW and Sec-FMCW waveforms as a function of the number of phase chips per chirp, $L$. It can be observed that the secure codebook design reduces the maximum throughput, since only a subset of the phase-codes is selected to satisfy the sensing-security constraints. This throughput reduction reflects the trade-off between communication rate and sensing privacy. Fig.~\ref{fig:secure_error} shows the bit error rates (BER) and packet error rates (PER) of the CU and C-Eve. It can be seen that the CU can demodulate all data and decode the message when sufficient SNR is available, achieving 0 BER and PER at around 10 dB SNR. On the other hand, C-Eve may demodulate IM data by using a maximum likelihood estimator, but it still has some errors even with high SNR. Moreover, C-Eve cannot demodulate the phase-coded data and decode the codebook due to the proposed two-layer modulation scheme, secure piloting, and lack of information about the codebook.

The physical-layer communication security performance in terms of throughput gap between the CU and C-Eve is shown in Fig.~\ref{fig:secure_rate}, where the throughput gap between CU and C-Eve can reach over 22 Mbits/s via the proposed receiver in Fig.~\ref{fig:Com_receiver}. In fact, C-Eve always has a very high block error rate, leading to incorrect demodulation of all data and zero throughput. Consequently, the CU's throughput is the same as the throughput gap shown in Fig.~\ref{fig:secure_rate}. 

Fig.~\ref{fig:range_profile} illustrates the target range profiles estimated by the receiver that employs a matched filter when there are three targets in the radar range. It can be seen that once the IM-FMCW (without phase coding) is used, the range profile of the targets does not have side-lobes, and using IM-PC-FMCW causes random sidelobes. However, the Sec-FMCW signaling introduces fixed high sidelobes, leading to ghost targets in the target range estimation. Having more phase transitions enables more flexible design of the AF as shown in Fig~\ref{fig:trade-off}, where the peak-sidelobe level (PSL) and integrated sidelobe level (ISL) of the signals are shown as a function of throughput. Higher PSL and ISL indicate that the signal has higher peak sidelobe levels, and this is achieved when the Sec-FMCW chirps have more phase segments. This also leads to higher data rates since more data can be delivered in each chirp due to the increasing phase code length. This increases the throughput gap between CU and C-Eve, leading to higher communication data security. 

Fig.~\ref{fig:range-velocity-eve} presents the range--velocity maps of the targets obtained by the ISAC BS and S-Eve, where both receivers operate at an SNR of 20 dB and integrate 64 chirps for sensing, and the reference link SNR at S-Eve is also 20 dB. Target velocity estimation requires coherent integration across multiple chirps; however, arbitrarily varying chirp parameters in the proposed secure signaling destroy this coherence for S-Eve. Since the ISAC sensing receiver knows the exact parameters of each transmitted chirp, it can compensate for the effects of varying bandwidth, center frequency, and phase coding by using the proposed radar processing algorithm in Section~\ref{sec:legitimate_radar}. As observed in Fig.~\ref{fig:range-velocity-eve}, this processing successfully restores the coherence across chirps and suppresses the sidelobes caused by these variations. In contrast, S-Eve does not know the exact chirp parameters and therefore cannot perform the corresponding compensation. As a result, its range--velocity maps exhibit strong sidelobes in both range and velocity when IM-PC-FMCW and Sec-FMCW waveforms are used, as shown in Fig.~\ref{fig:range-velocity-eve}. These sidelobes are significantly stronger for Sec-FMCW, leading to ghost targets and large range and velocity estimation errors. Therefore, the proposed secure signaling prevents S-Eve from accurately estimating the target parameters, thereby significantly improving sensing privacy while also enhancing communication security.

Fig.~\ref{fig:error_stats} shows the RMSE of the range and velocity estimates obtained by the ISAC BS and S-Eve, where both receivers employ cell-averaging constant false alarm rate (CA-CFAR) detection with a false alarm probability of \(10^{-6}\). The considered scenario is the same as that in Fig.~\ref{fig:range-velocity-eve}, and the RMSE values are evaluated for the target located at 100 m. Fig.~\ref{fig:error_stats}(a) shows that S-Eve requires a substantially higher SNR than the ISAC BS for accurate range estimation (around 20 dB SNR gap), because its estimation process is directly affected by the randomly varying chirp parameters, including bandwidth, center frequency, and the phase codes. Moreover, when Sec-FMCW is employed, the range RMSE of S-Eve exhibits a clear error floor, indicating that S-Eve cannot achieve accurate range estimation even at high SNR. Velocity estimation is even more challenging for S-Eve, as shown in Fig.~\ref{fig:error_stats}(b). The arbitrarily varying chirp parameters introduce irregular phase variations across chirps as seen in Fig.~\ref{fig:range-velocity-eve}, which destroy the coherence required for Doppler estimation. The ISAC BS can compensate for these phase variations because it has full knowledge of the chirp parameters, and thus it can estimate the target velocity accurately using the proposed radar receiver algorithm when the SNR is sufficiently high. In contrast, S-Eve cannot estimate the target velocity even at very high SNR due to the very high Doppler ambiguity caused by varying chirp parameters. These results demonstrate that the proposed secure ISAC waveform design substantially degrades the sensing capability of S-Eve.

As shown in the numerical results, the proposed radar-centric secure ISAC signaling delivers robust physical-layer security by simultaneously protecting communication data and degrading unauthorized sensing capabilities. By utilizing a two-layer modulation scheme and secure piloting, the system ensures high throughput and secure communication for legitimate users while preventing C-Eve from successfully decoding the data. Furthermore, the deliberate use of randomly varying chirp parameters destroys the phase coherence required for Doppler estimation, causing S-Eve to completely fail at velocity tracking and suffer severe error floors in range estimation. Because the legitimate ISAC base station retains full knowledge of these parameters, it can compensate for the phase variations to maintain accurate target tracking.

\section{Conclusion}
This study presents a radar-centric secure ISAC signal design and transceiver architecture that provides high physical-layer security for both communication and sensing. By utilizing a combination of varying chirp parameters, AF optimization, and codebook design, the proposed approach enables highly secure physical-layer data transmission while deteriorating the passive sensing capabilities of sensing eavesdroppers. Varying the chirp bandwidth and center frequency introduces arbitrary distortions in the range-velocity maps obtained by S-Eve; however, these distortions can be compensated at the legitimate receiver using the proposed algorithm. In addition, AF design aims to generate ghost targets in the range images obtained by S-Eve, thereby degrading its target range estimation capability. Meanwhile, the proposed ISAC signaling and codebook design, together with the two-layer modulation scheme, enhance physical-layer communication security, as both index modulation and phase coding must be successfully demodulated to decode the transmitted data.

\section*{Acknowledgment}
This study is supported in parts by the  Advanced Research Next Generation Information Networks (AR-NGIN) project of the DSTL, by the Smart Networks and
Services Joint Undertaking (SNS JU) project 6G-MUSICAL under Grant Agreement No. 101139176, and by the Engineering and Physical Sciences Research Council UK-India project
ICON with Grant Agreement No UKRI859.

\ifCLASSOPTIONcaptionsoff
  \newpage
\fi

\bibliographystyle{IEEEtran}
\bibliography{Ref.bib}

\end{document}